\documentclass[aps,twocolumn,showpacs,preprintnumbers,nofootinbib,prl,superscriptaddress,groupedaddress,10pt]{revtex4-1}
\makeatletter
\def\l@subsubsection#1#2{}
\def\l@subsubsubsection#1#2{}
\makeatother

\usepackage{graphicx,amssymb,amsmath,amsthm,amsfonts,epsfig,epsf}
\usepackage[usenames]{color}
\usepackage{epstopdf}
\definecolor{darkred}{rgb}{0.5,0,0}
\usepackage{bm}
\usepackage{dcolumn}
\usepackage[latin1]{inputenc}
\usepackage{latexsym}
\usepackage{rotating}
\usepackage{longtable}

\usepackage{enumerate}
\usepackage{tensor,multirow}
\usepackage{url,pifont}

\usepackage{gensymb}

\usepackage{ulem}

\def\be{\begin{equation}}
\def\ee{\end{equation}}
\newcommand{\beq}{\begin{eqnarray}}
\newcommand{\eeq}{\end{eqnarray}}

\def\ba{\begin{align}}
\def\ea{\end{align}}

\begin{document}

\title{Self-gravitating oscillons and new critical behavior}

\author{Taishi Ikeda$^{1}$\footnote{Electronic address: ikeda@gravity.phys.nagoya-u.ac.jp},
Chul-Moon~Yoo$^{1}$\footnote{Electronic address: yoo@gravity.phys.nagoya-u.ac.jp},
Vitor Cardoso$^{2,3}$\footnote{Electronic address: vitor.cardoso@ist.utl.pt}
}
%
%\affiliation{${^1}$ Departure \vc{Department?} of Physics,~~Graduate School of Science,~~Nagoya University,~~Nagoya~~464-6602,~~Japan}
\affiliation{${^1}$ Department of Physics,~~Graduate School of Science,~~Nagoya University,~~Nagoya~~464-6602,~~Japan}
\affiliation{${^2}$ CENTRA, Departamento de F\'{\i}sica, Instituto Superior T\'ecnico, Universidade de Lisboa, Avenida~Rovisco Pais 1, 1049 Lisboa, Portugal}
\affiliation{${^3}$ Perimeter Institute for Theoretical Physics, 31 Caroline Street North Waterloo, Ontario N2L 2Y5, Canada}

\begin{abstract}
The dynamical evolution of self-interacting scalars is of paramount importance in cosmological settings, and can teach us about the
content of Einstein's equations. In flat space, nonlinear scalar field theories can give rise to
localized, non-singular, time-dependent, long-lived solutions called {\it oscillons}. 
Here, we discuss the effects of gravity on the properties and formation of these structures,
described by a scalar field with a double well potential.
We show that oscillons continue to exist even when gravity is turned on, and we conjecture that
there exists a sequence of critical solutions with infinite lifetime.
Our results suggest that a new type of critical behavior appears in this theory, characterized by
modulations of the lifetime of the oscillon around the scaling law and the modulations of the amplitude of the critical solutions.
\end{abstract}

% \pacs{04.70.-s,04.30.-w,04.30.Tv}

\maketitle

%%%%%%%%%%%%%%%%%%%%%%%%%%%%%%%%%%%%%%%%%%%%%%
\section{Introduction}\label{Sec-introduction}
%%%%%%%%%%%%%%%%%%%%%%%%%%%%%%%%%%%%%%%%%%%%%%
Our understanding of the physical content and intricate dynamics of
non-linear field equations is in its infancy. The understanding of universal features associated with non-linear dynamics is one of the lines of investigation pursued more vigorously.
The underlying motivation is that, even in the absence of precise predictions, universal features are a useful guide to the understanding
of the qualitative behavior of any system governed by such dynamics. The study of universal phenomena in non-linear systems may provide not only new insight into complex systems, but also clues to solve outstanding problems. One of the most important and difficult problems in physics concerns the existence of solitonic solutions
of the field equations, their formation, evolution and stability. While a general understanding of this problem is missing, there were important developments
in the last few decades. The nonlinear stability of Minkowski spacetime was established rigorously decades ago for massless fields~\cite{Christodoulou} and more recently for massive scalars~\cite{Okawa:2013jba,LeFloch:2015ppi}: arbitrarily small initial perturbations eventually disperse to infinity. As the amplitude of the initial
data is tuned up, collapse to a black hole eventually ensues, driven by nonlinear gravitational effects~\cite{Choptuik:1992jv}. Interestingly, the development
of initial data close to the critical solution displays universality~\cite{comment:2017}. Thus, despite the complicated dynamics associated to Einstein equations,
and the violent character of gravitational collapse, at least some features of the process are universal~\cite{Choptuik:1992jv,Gundlach:2007gc}.

%Studies of gravitational collapse were rapidly extended to other theories and setups.
Here, we focus on an Einstein-scalar system with a double well potential of the form
\begin{equation}
V(\Phi)=\frac{\lambda}{4}(\Phi^{2}-\sigma^{2})^{2}. \label{eq-potential}
\end{equation}
The theory, interesting from a cosmological and particle physics perspective, contains two distinct sources of non-linearity. One is the non-linear gravitational interaction of Einstein gravity, the other is non-linear self interaction of the scalar field.
The non-linear effects of the double well potential have been actively analyzed in a Minkowski background. 
It was found that, in such potentials, the theory has localized dynamical long-lived solutions, which are called ``oscillons''~\cite{Bogolyubsky:1976nx,Copeland:1995fq,Honda:2001xg,Fodor:2006zs,Gleiser:2004an}.

In a Minkowski background, oscillons were firstly observed by Bogolyubsky et al~\cite{Bogolyubsky:1976nx}, and subsequently
studied in more depth by Copeland et al~\cite{Copeland:1995fq}. It was found that oscillons can be the result of collapse of a vacuum region surrounded by a domain wall.
The lifetime $\tau$ of the oscillon, which depends on the initial bubble radius, is much larger than 
the typical dynamical scale of the system. The typical energy of a oscillon is about $\frac{43}{\sigma}$~~\cite{Copeland:1995fq}.
Therefore, if the initial energy of the bubble is smaller than this number, the oscillon is not able to form, and the scalar field disperses away.
Subsequently, it was pointed out that the lifetime of oscillons depends non-trivially on initial conditions~\cite{Honda:2001xg, Fodor:2006zs}. 
In particular, oscillons can become infinitely long-lived if the initial bubble radius $R_{0}$ is fine-tuned to some value $R_{\ast}$.
Around $R_{\ast}$, $\tau$ obeys the scaling law: 
\be
\tau =-\gamma\ln |R-R_{\ast}|+C\,,\label{scaling_law}
\ee
where $C$ and $\gamma$ are constants. This behavior is similar to type I critical collapse~\cite{Gundlach:2007gc}.
Our goal is to understand the structure of oscillons and critical collapse in this theory, once gravity is turned on.
As a first step, in this paper, we analyze effects of gravity on the oscillons in the relatively weak gravity case.

This paper is organized as follows. In Sec. \ref{Sec-formulation}, we explain our settings and numerical procedure.  
We also introduce the definitions of the mass and the lifetime of an oscillon in Sec.\ref{Sec-formulation}. 
Results are shown in Sec.\ref{Sec-Result}, and Sec. \ref{Sec-Summary and discussion} 
is devoted to a summary and discussion. We use the units in which the speed of light is equal to unity.

%%%%%%%%%%%%%%%%%%%%%%%%%%%%%%%%%%%%%%%%%%%
\section{Setting and Numerical Formulation}
\label{Sec-formulation}
%%%%%%%%%%%%%%%%%%%%%%%%%%%%%%%%%%%%%%%%%%%

%%%%%%%%%%%%%%%%%%%%%%%%%%%%%%%%%%%
\subsection{Einstein-scalar system}
%%%%%%%%%%%%%%%%%%%%%%%%%%%%%%%%%%%
We consider the Einstein-scalar theory described by the action
\begin{eqnarray}\label{eq-action}
S=\int d^4x \sqrt{-g}\left(\frac{R}{8\pi G}-g^{\mu\nu}\Phi_{,\mu}\Phi_{,\nu}-2V(\Phi)\right)\,.
\end{eqnarray}
The corresponding equations of motion are
\begin{eqnarray}\label{eq-lagrangian}
G_{\mu\nu}&=&8\pi G \left(-\frac{1}{2}g_{\mu\nu}(\nabla\Phi)^{2}+\nabla_{\mu}\Phi\nabla_{\nu}\Phi-g_{\mu\nu}V(\Phi)\right)\,,\\
\nabla^{2}\Phi&=&V^{\prime}(\Phi)\,,
\end{eqnarray}
where $G$ is Newton's constant, $g_{\mu\nu}$ is the spacetime metric and $G_{\mu\nu}$ is the Einstein tensor associated with $g_{\mu\nu}$. 
$\Phi$ is a scalar field and $V(\Phi)$ is the double well potential given by \eqref{eq-potential}.

In this system, a typical length scale is given by $1/\sqrt{\lambda \sigma^2}$, 
and we use this combination as the length scale unit 
\be
L\equiv 1/\sqrt{\lambda \sigma^2}\,,
\ee
throughout this paper. 
The dimensionless combination $G\sigma^2$ can be used to characterize 
the strength of the gravitational interaction~\footnote{
Introducing a typical length $L$, from Eqs.(\ref{eq-potential}, \ref{eq-lagrangian}), 
we can get the following equations about the units: 
$1/[L]^{2}=[G][\Phi]^{2}/[L]^{2}=[\lambda][\Phi]^{4}=[\lambda][\sigma]^{4}$.
Therefore, a set of dimensionless combinations of the variables are $\Phi/\sigma$, $G\sigma^{2}$ and $L\sqrt{\lambda\sigma^{2}}$.
.}.

%%%%%%%%%%%%%%%%%%%%%%%%%%%%%%%%%%%%%%%%%%%%%%%%%%%%%
\subsection{Formulation and numerical implementation}
%%%%%%%%%%%%%%%%%%%%%%%%%%%%%%%%%%%%%%%%%%%%%%%%%%%%%
We will solve the above set of nonlinear equations numerically.
Since oscillons are long-lived solutions, one needs to perform accurate long-term numerical simulations.
In order to achieve a long-term accurate numerical simulations of the Einstein equations, 
we adopt the free evolution scheme of the generalized Baumgarte-Shapiro-Shibata-Nakamura (G-BSSN) formulation. 
The G-BSSN formulation is a generalization of the BSSN formulation~\cite{Shibata:1995we, Baumgarte:1998te} to the case of curvilinear coordinates\cite{Brown:2009dd, Alcubierre:2010is},
and is useful for time evolutions of spherically symmetric spacetimes~\cite{Akbarian:2015oaa}. 
The general version of G-BSSN formulation is presented in Appendix \ref{Sec.G-BSSN formulation} for completeness.
First, let us summarize the G-BSSN formulation in spherically symmetric spacetimes 
and our specific procedures to solve the time evolution.
%%%%%%%%%%%%%%%%%%%%%%%%%%%%%%%%%%%%%%%%%%%%%%%%%%%%%%%%%%%%%%%%%%%%%%
\subsubsection{G-BSSN formulation in spherically symmetric spacetime}
%%%%%%%%%%%%%%%%%%%%%%%%%%%%%%%%%%%%%%%%%%%%%%%%%%%%%%%%%%%%%%%%%%%%%%
Let us start with the following general expression of the line element:
\begin{equation}
ds^{2}=-\alpha^{2}dt^{2}+\gamma_{ij}(dx^{i}+\beta^{i}dt)(dx^{j}+\beta^{j}dt),
\end{equation}
where 
$\alpha$, $\beta^i$ and $\gamma_{ij}$ are the lapse function, shift vector and spatial 3-metric, respectively.
Under the spherical symmetry assumption, and 
introducing the spherical coordinates $(r,\theta,\phi)$, 
we can write the coordinate basis components of $\tilde{\gamma}_{ij}$ and $\tilde{A}_{ij}$, which are defined in Eqs.(\ref{def tilde gamma}) (\ref{def tilde A K}), as follows: 
$\tilde{\gamma}_{ij}=\mbox{diag}(a,br^{2},br^{2}\sin^{2}\theta)$ and $\tilde{A}_{ij}=\mbox{diag}(A,Br^{2},Br^{2}\sin^{2}\theta)$, 
where $a$, $b$, $A$ and $B$ are functions of $t$ and $r$. 
Furthermore, because of spherical symmetry, 
$\beta^{i}=(\beta,0,0)$ and $\tilde{\Lambda}^{i}=(\tilde{\Lambda},0,0)$, 
where $\beta$ and $\tilde \Lambda$ are functions of $t$ and $r$, 
and the definition of $\tilde \Lambda^i$ is given in Appendix \ref{Sec.G-BSSN formulation}.

Using this coordinate system, 
the equations of motion for the scalar field $\Phi$ and its conjugate momentum $\Pi$ can be explicitly written as follows:
\begin{eqnarray}
\partial_{t}\Pi&=&\beta\Pi^{\prime}+\left(\frac{2}{3}\alpha K+2\alpha\frac{B}{b}+\beta^{\prime}\right)\Pi\nonumber\\
&&+\left(\frac{\alpha^{\prime}}{e^{2\phi}\sqrt{a}}+\frac{2\alpha\phi^{\prime}}{e^{2\phi}\sqrt{a}}-\frac{\alpha}{2e^{2\phi}\sqrt{a^{3}}}a^{\prime}+ \frac{\alpha}{e^{2\phi}\sqrt{a}}\frac{b^{\prime}}{b}\right.\nonumber\\
&&\left.+ \frac{2\alpha}{re^{2\phi}\sqrt{a}}\right)\Phi^{\prime}+\frac{\alpha}{e^{2\phi}\sqrt{a}}\Phi^{\prime\prime}-\alpha e^{2\phi}\sqrt{a}\frac{dV}{d\Phi}%(\Phi)
,\\
\partial_{t}\Phi&=&\beta\Phi^{\prime}+\frac{\alpha}{e^{2\phi}\sqrt{a}}\Pi,
\end{eqnarray}
where the prime denotes a derivative with respect to $r$. 
Non-zero components of the energy momentum tensor are the energy density $E$, radial component of the momentum $p$ and three components of 
stress tensor $S_{rr}$, $S_{\theta\theta}$ and $S_{\phi\phi}$ defined as follows:
\begin{eqnarray}
E&:= &T_{\mu\nu}n^{\mu}n^{\nu}=e^{-4\phi}\frac{\Pi^{2}+\Phi^{\prime 2}}{2a}+V,\\
p&:=&-T_{\nu\mu}\gamma^{\nu}_{r}n^{\mu}=-\frac{\Pi\Phi^{\prime}}{e^{2\phi}\sqrt{a}},\\
S_{rr}&:=&T_{\mu\nu}\gamma^{\mu}_{r}\gamma^{\nu}_{r}=\frac{\Pi^{2}}{2}+\frac{\Phi^{\prime 2}}{2}-e^{4\phi}aV,\\
S_{\theta\theta}&:=&T_{\mu\nu}\gamma^{\mu}_{\theta}\gamma^{\nu}_{\theta}=
-\frac{br^{2}}{2a}(-\Pi^{2}+\Phi^{\prime 2})-e^{4\phi}br^{2}V,\\
S_{\phi\phi}&:=&T_{\mu\nu}\gamma^{\mu}_{\phi}\gamma^{\nu}_{\phi}=S_{\theta\theta}\sin^{2}\theta.
\end{eqnarray}

The evolution equations for the geometrical variables $\alpha$, $\beta$, $\phi$, $a$, $b$, $A$, $B$, $K$, $\tilde{\Lambda}$ are given as follows:
\begin{widetext}
\begin{eqnarray}
\partial_{t}\phi&=&\beta\phi^{\prime}-\frac{1}{6}\alpha K+\kappa\frac{1}{6}\mathcal{B},\\
\partial_{t}a&=&\beta a^{\prime}+2a\beta^{\prime}-2\alpha A -\kappa\frac{2}{3}a\mathcal{B},\\
\partial_{t}b&=&\beta b^{\prime}+2\beta\frac{b}{r}-2\alpha B -\kappa\frac{2}{3}b\mathcal{B},\\
\partial_{t}K&=&\beta K^{\prime}-\mathcal{D}+\alpha\left(\frac{1}{3}K^{2}+\frac{A^{2}}{a^{2}}+2\frac{B^{2}}{b^{2}}\right)+4\pi\alpha(E+S),\\
\partial_{t}A&=&\beta A^{\prime}+2A\beta^{\prime}+e^{-4\phi}\left\{-\mathcal{D}_{rr}^{TF}+\alpha(R_{rr}^{TF}-8\pi S_{rr}^{TF})\right\}+\alpha \left(KA-2\frac{A^{2}}{a}\right)-\kappa\frac{2}{3}A\mathcal{B},\nonumber\\
\\
\partial_{t}B&=&\beta B^{\prime}+\frac{e^{-4\phi}}{r^{2}}\left\{-\mathcal{D}_{\theta\theta}^{TF}+\alpha(R_{\theta\theta}^{TF}-8\pi S_{\theta\theta}^{TF})\right\}+\alpha\left(KB-2\frac{B^{2}}{b}\right)+2\frac{\beta}{r}B-\kappa\frac{2}{3}B\mathcal{B},\nonumber\\
\\
\partial_{t}\tilde{\Lambda}&=&\beta\tilde{\Lambda}^{\prime}-\beta^{\prime}\tilde{\Lambda}+\frac{2\alpha}{a}\left(\frac{6 A\phi^{\prime}}{a}-\frac{2}{3}K^{\prime}-8\pi S_{r}\right)
+\frac{\alpha}{a}\left(\frac{a^{\prime}A}{a^{2}}-\frac{2b^{\prime}B}{b^{2}}+4B\frac{a-b}{rb^{2}}\right)\nonumber\\
&&+\kappa\left(\frac{2}{3}\tilde{\Lambda}\mathcal{B}+\frac{\mathcal{B}^{\prime}}{3a}\right)+\frac{2}{rb}\left(\beta^{\prime}-\frac{\beta}{r}\right)-2\frac{\alpha^{\prime}A}{a^{2}}+\frac{1}{a}\beta^{\prime\prime},
\end{eqnarray}
\end{widetext}
where $R_{rr}$ and $R_{\theta\theta}$ are each component of 
the spatial Ricci tensor, and 
$\mathcal{D}_{ij}$, $\mathcal{D}$ and $\mathcal B$ are defined by 
$\mathcal{D}_{ij}:= D_{i}D_{j}\alpha$, $\mathcal{D}:=\gamma^{ij}D_{i}D_{j}\alpha$ 
and $\mathcal{B}:= \tilde{D}_{k}\beta^{k}$ respectively. 
The superscript ${\rm TF}$ denotes the trace-free part of the tensor.
The explicit expressions of $\mathcal{B},~\mathcal{D}_{rr},~\mathcal{D}_{\theta\theta},~R_{rr},R_{\theta\theta}$
are given in Appendix~\ref{expressions}. 
Here, we have chosen the Lagrangian type for the time evolution of $\mbox{det}\tilde{\gamma}$, that is,
$\partial_t\tilde\gamma=0$ and $\kappa=1$ (see Appendix~\ref{Sec.G-BSSN formulation}). 

The Hamiltonian and momentum constraints are expressed as follows:
\begin{eqnarray}
&&\left\{\frac{\phi^{\prime\prime}}{a}+\frac{\phi^{\prime 2}}{a}-(\frac{a^{\prime}}{2a^{2}}-\frac{b^{\prime}}{ab}-\frac{2}{ar})\phi^{\prime}\right\}e^{\phi}-\frac{e^{\phi}}{8}\tilde{R}\nonumber\\
&+&\frac{e^{5\phi}}{8}\left(\frac{A^{2}}{a^{2}}+2\frac{B^{2}}{b^{2}}\right)-\frac{e^{5\phi}}{12}K^{2}+2\pi e^{5\phi}E=0,\nonumber\\
&&6\phi^{\prime}\frac{A}{a}+\frac{A^{\prime}}{a}-\frac{a^{\prime}A}{a^{2}}+\frac{b^{\prime}}{b}\left(\frac{A}{a}-\frac{B}{b}\right)+\frac{2}{r}\left(\frac{A}{a}-\frac{B}{b}\right)\nonumber \\
&-&\frac{2}{3}K^{\prime}-8\pi p=0\,.
\end{eqnarray}
%
%%%%%%%%%%%%%%%%%%%%%%%%%%%%%%%%%
\subsubsection{Gauge conditions}
%%%%%%%%%%%%%%%%%%%%%%%%%%%%%%%%%
It is necessary to choose a gauge condition for the lapse function $\alpha$ and the shift vector $\beta$, respectively.
For the lapse function,
we choose the harmonic gauge:
\begin{equation}
\partial _{t}\alpha=\mathcal{L}_{\beta}\alpha-K\alpha^{2},
\end{equation}
and for the shift vector,
we employ the normal coordinate: $\beta=0$.
%%%%%%%%%%%%%%%%%%%%%%%%%%%%%%%%%%%%%
\subsubsection{Boundary conditions}
%%%%%%%%%%%%%%%%%%%%%%%%%%%%%%%%%%%%
First, we require the regularity for the derivative of every variable as follows:
\begin{eqnarray}
\alpha^{\prime}|_{t=0}&=&\phi^{\prime}|_{r=0}=K^{\prime}|_{r=0}=\tilde{\Lambda}|_{r=0}\nonumber\\
&=&\Phi^{\prime}|_{r=0}=\Pi^{\prime}|_{r=0}=0,\label{boundary-regularity-other}
\\
a^{\prime}|_{r=0}&=&b^{\prime}|_{r=0}=A^{\prime}|_{r=0}=B^{\prime}|_{r=0}=0. 
\label{boundary-regularity-neuman}
\end{eqnarray}
On the other hand,
the following second boundary conditions are imposed for $a$, $b$, $A$ and $B$ 
by requirements that the evolution equations must be regular at the origin \cite{Alcubierre:2004gn},
\begin{equation}
a|_{r=0}-1=b|_{r=0}-1=
A|_{r=0}=B|_{r=0}=0. \label{boundary-local-flat}
\end{equation}
Through these conditions, the
right-hand side of the evolution equations are regularized.
If the boundary conditions Eq.(\ref{boundary-local-flat}) are satisfied for an initial data set, 
the evolution equations guarantee that these boundary conditions are always satisfied.
However,
if we numerically solve the evolution equations,
these conditions can be violated within numerical precision and may trigger numerical instabilities.
In this paper, in order to avoid numerical instabilities,
we use Eqs.(\ref{boundary-regularity-other}) and (\ref{boundary-local-flat}) \cite{Akbarian:2015oaa}. 
In addition, we evaluate the values of $\alpha$, $\phi$, $K$,$\Phi$ and $\Pi$ 
at the first outer grid from the origin, using \eqref{boundary-regularity-other} combined with 
forward finite differencing~\footnote
{If $f(r)$ is a even function of $r$ at the origin,
boundary condition for $f(r)$ is as follows:$f^{\prime}|_{r=0}=0~(~^{\forall}t)$.
Therefore,
its time derivative $\partial_{t}f^{\prime}|_{r=0}=0$ must vanish.
By using the forward finite differencing,
we get the following formula:
$f_{1}^{n+1}=\frac{1}{3}(4f_{2}^{n+1}-f_{3}^{n+1}-3f_{1}^{n}+4f_{2}^{n}-f_{3}^{n})$
where $f^{n}_{i}$ denotes the value at the $n$'s time step and $i$'s grid number.
The grid $i=0$ corresponds to $r=0$.}.

The outer boundary conditions are described as follows:
\begin{eqnarray}
\phi(t,r_{\mbox{\footnotesize{max}}})&=&\log \left(1+\frac{M_{\mbox{\footnotesize{ADM}}}}{2r_{\mbox{\footnotesize{max}}}}\right),\label{outer-boundary-condition-phi}\\
\alpha(t,r_{\mbox{\footnotesize{max}}})&=&a(t,r_{\mbox{\footnotesize{max}}})=b(t,r_{\mbox{\footnotesize{max}}})=1,
\label{outer-boundary-condition-unity}\\
A(t,r_{\mbox{\footnotesize{max}}})&=&B(t,r_{\mbox{\footnotesize{max}}})=K(t,r_{\mbox{\footnotesize{max}}})\nonumber\\
&=&\tilde{\Lambda}(t,r_{\mbox{\footnotesize{max}}})=\Pi(t,r_{\mbox{\footnotesize{max}}})=0,\label{outer-boundary-condition-zero}\\
\Phi(t,r_{\mbox{\footnotesize{max}}})&=&\sigma,\label{outer-boundary-condition-Phi}
\end{eqnarray}
where $r_{\mbox{\footnotesize{max}}}$ is
the coordinate value at the outer boundary and
$M_{\mbox{\footnotesize{ADM}}}$ is the ADM mass.
%%%%%%%%%%%%%%%%%%%%%%%%%%%%%%%%%%%%%%%%%%%%%%%%%%%%%%%%
\subsubsection{Inhomogeneous grid}
\begin{figure}
\begin{tabular}{cc}
\includegraphics[scale=0.5,clip]{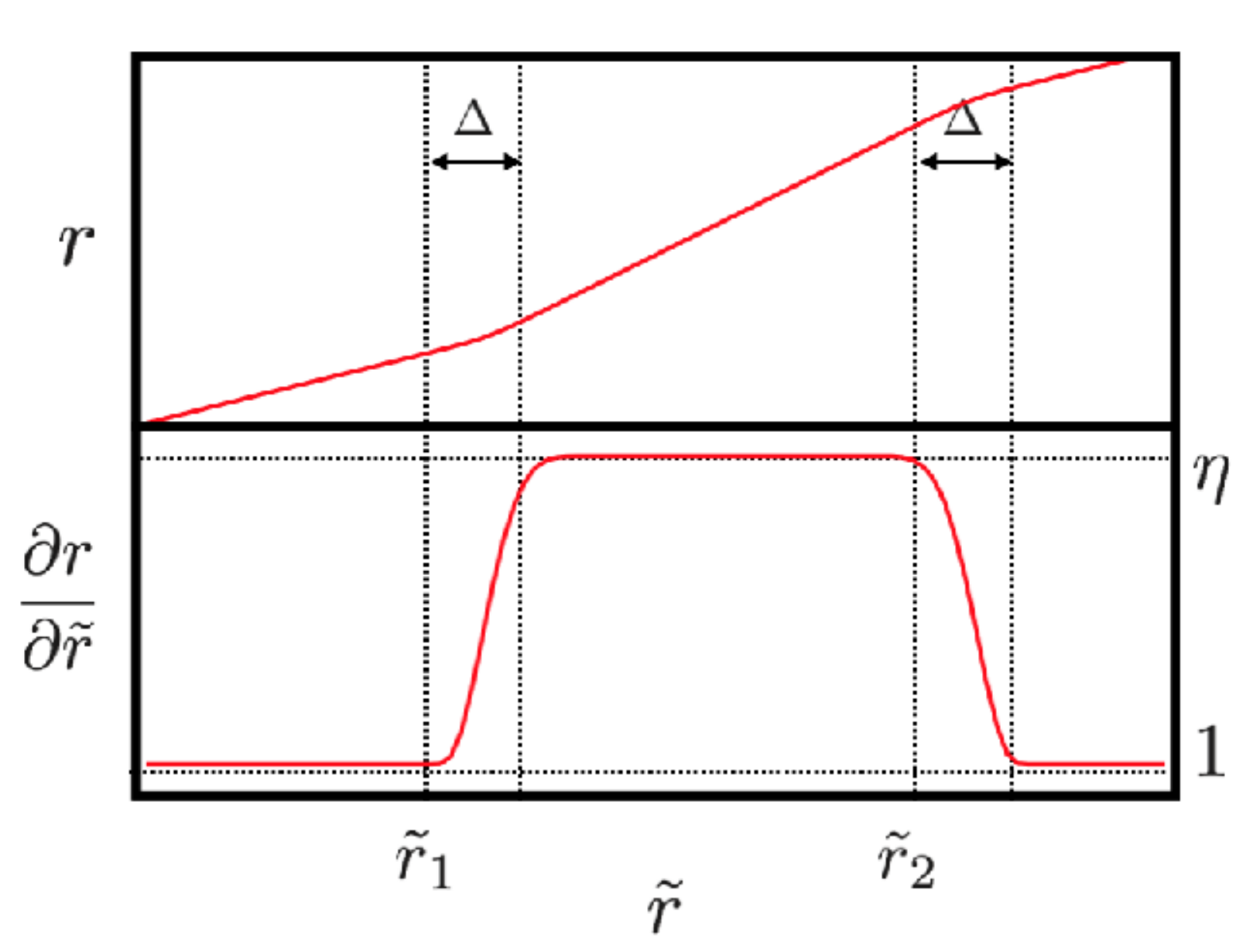}
\end{tabular}
\caption{\label{Fig-inhom1}
$r$ and $\partial r/\partial \tilde r$ are schematically depicted 
as functions of $\tilde r$. 
}
\end{figure}

%%%%%%%%%%%%%%%%%%%%%%%%%%%%%%%%%%%%%%%%%%%%%%%%%%%%%%%%
In order to obtain an accurate long-term numerical evolution, unphysical reflections of the scalar at the outer boundary
need to be avoided. One of the simplest ways to accomplish this is to place the numerical boundary 
far away, by making the proper distance large enough~\cite{Copeland:1995fq, Honda:2001xg}.
Here, we introduce an inhomogeneous grid spacing by performing the following radial coordinate transformation:
\begin{widetext}
\begin{equation}\label{coordinate transformation}
\frac{\partial r}{\partial \tilde{r}}=
\left\{
\begin{array}{lc}
1&(0<\tilde{r}<\tilde{r}_{1})\\
1+(-1+\eta)\left\{ \Delta^{4}-(\tilde{r}_{1}+\Delta-\tilde{r})^{4}\right\}^{4}\Delta^{-16}&(\tilde{r}_{1}<\tilde{r}<\tilde{r}_{1}+\Delta)\\
\eta&(\tilde{r}_{1}+\Delta<\tilde{r}<\tilde{r}_{2})\\
1+(-1+\eta)\left\{\Delta^{4}-(\tilde{r}_{2}-\tilde{r})^{4}\right\}^{4}\Delta^{-16}&(\tilde{r}_{2}<\tilde{r}<\tilde{r}_{2}+\Delta)\\
1&(\tilde{r}_{2}+\Delta<\tilde{r}),
\end{array}
\right.
\end{equation}
\end{widetext}
where $\tilde{r}$ is the new radial coordinate,
and $\Delta$ ,$\eta$ $\tilde{r}_{1}$ and $\tilde{r}_{2}$ are parameters of the inhomogeneous grid spacing (See Fig.~\ref{Fig-inhom1}).
We set these parameters as follows:
\begin{widetext}
\begin{eqnarray}
(\Delta/L~,\eta~,\tilde{r}_{1}/L,\tilde{r}_{2}/L)&=(10,1.25,80,\tilde{r}_{\footnotesize{\mbox{max}}}-20)~\mbox{for}~\Delta \tilde{r}/L=2.0\times 10^{-2},\\
(\Delta/L~,\eta~,\tilde{r}_{1}/L,\tilde{r}_{2}/L)&=(10,2.50,80,\tilde{r}_{\footnotesize{\mbox{max}}}-20)~\mbox{for}~\Delta \tilde{r}/L=1.0\times 10^{-2},
\end{eqnarray}
\end{widetext}
where $\tilde{r}_{\footnotesize{\mbox{max}}}$ is a numerical boundary of the new radial coordinate,
and it is chosen so that the areal radius of the numerical boundary is farther than the oscillon's lifetime (i.e, the boundary is causally disconnected from the evolution).
Accordingly, 
we use the boundary conditions which are given 
from the coordinate transformation Eq.~(\ref{coordinate transformation}) of 
Eqs.~(\ref{outer-boundary-condition-phi})-(\ref{outer-boundary-condition-Phi}).
%%%%%%%%%%%%%%%%%%%%%%%%%%%%%%%%%%%%%%%%%%%%%%%%%%%
\subsection{Definitions of the mass and lifetime}
\subsubsection{Kodama mass}
%%%%%%%%%%%%%%%%%%%%%%%%%%%%%%%%%%%%%%%%%%%%%%%%%%%
%
\begin{figure*}[th]
\begin{tabular}{cc}
\includegraphics[scale=0.65,clip]{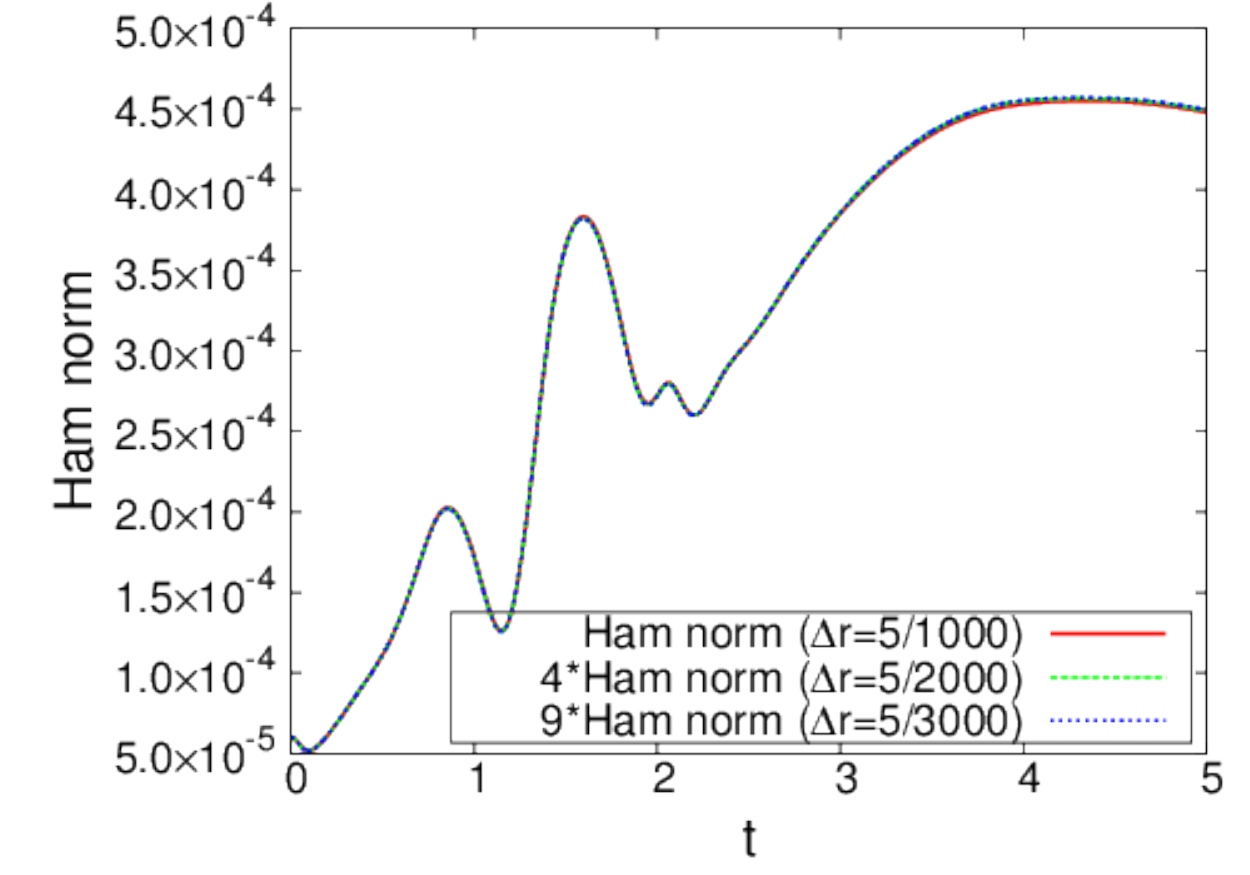}&
\includegraphics[scale=0.65,clip]{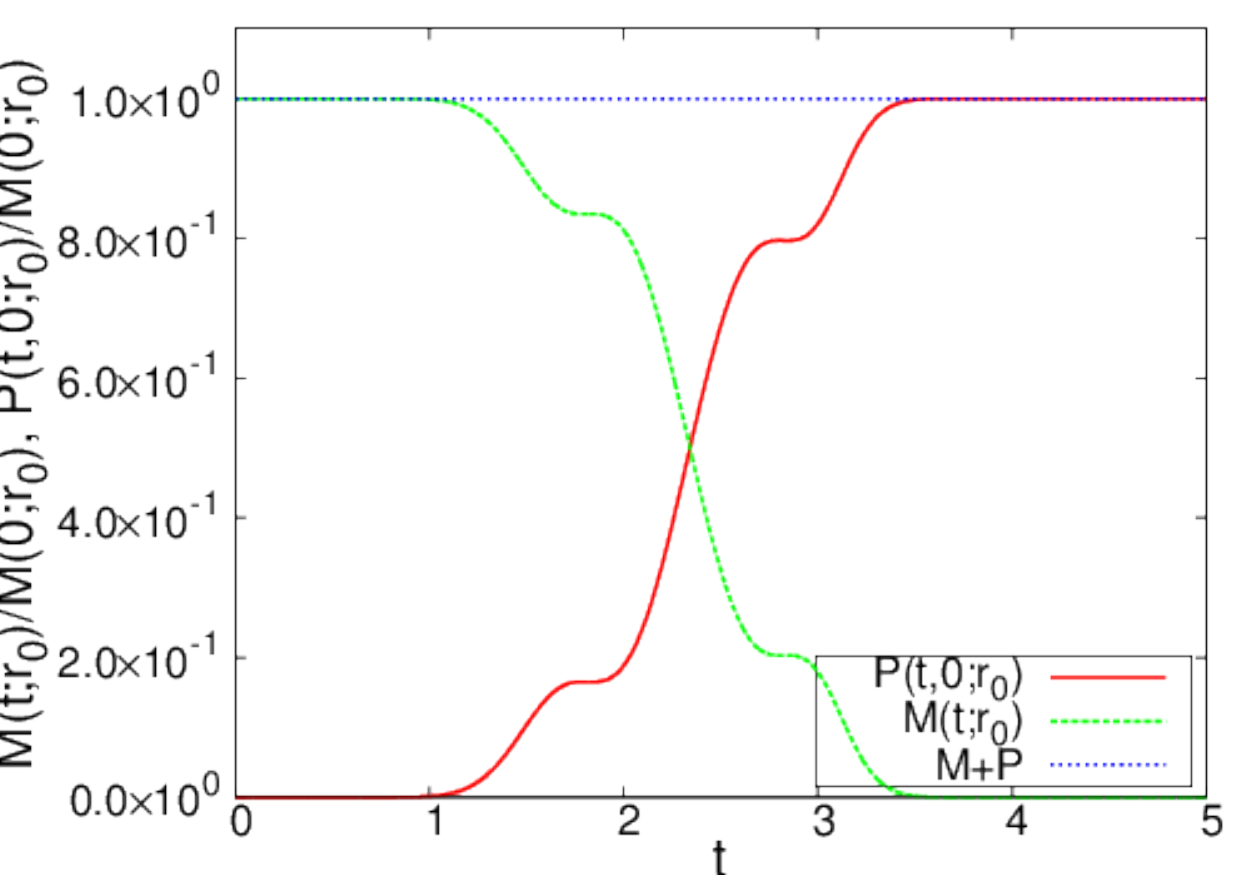}
\end{tabular}
\caption{\label{Fig.graph_convergence_ham_norm}
Left Panel: the L2 norm of Hamiltonian constraint violation, for different grid resolutions.
We have re-scaled the line in a way appropriate for second-order accurate codes. The near perfect overlap shows that indeed our results 
display second order convergence. Right Panel: Study of the conservation of the Kodama mass. 
}
\end{figure*}
The lifetime of these structures is an ambiguous concept. To make a precise definition, we will first define the concept of Kodama mass~\cite{Kodama:1979vn, Harada:2015yda}, 
which is a quasi-local conserved energy in a spherically symmetric spacetime. 
The Kodama mass is defined by using the Kodama vector $K^{\mu}$. 
Consider the two dimensional manifold
which is embedded as a two-dimensional surface of constant angular coordinates in a four dimensional manifold. 
We use the time coordinate $t$ and the radial coordinate $r$ to span this two-dimensional surface. 
By using the induced metric $G_{AB}$, 
the Kodama mass in this two dimensional manifold is defined as follows:
\begin{equation}
K^{A}=\epsilon^{AB}\partial_{B}R,
\end{equation}
where $R$ is the areal radius of the 2-sphere with constant $t$ and $r$. 
$\epsilon_{AB}$ is the Levi-Civita tensor on the 2-sphere. 
%given by $\epsilon_{AB}=\sqrt{-\mbox{det}(G)}\varepsilon_{AB}$ with the Levi-Civita tensor $\varepsilon_{AB}$.
We can naturally extend $K^{A}$ on the two dimensional manifold 
to $K^{\mu}$ on the four dimensional manifold. 
Then,
we define the four vector $S^{\mu}$ as follows:
\begin{equation}
S^{\mu}=T^{\mu\nu}K_{\nu}\,,
\end{equation}
where $T^{\mu\nu}$ is the energy momentum tensor. 
$S^{\mu}$ is the conserved four flux and 
satisfies the following conservation law:
\begin{equation}
\partial_{\mu}(\sqrt{-g}S^{\mu})=0\,.
\end{equation}
Therefore, we can define the conserved mass $M$, so-called Kodama mass, 
in a sphere of the radius $r_{0}$ on a constant $t$ hyper-surface as follows: 
\begin{equation}
M(t,r_{0})=\int_{\mbox{\footnotesize{sphere}}} S^{t}\alpha\sqrt{\gamma}dx^{3}. 
\end{equation}
In our case,
the Kodama mass can be written as follows:
\begin{equation}
M(t;r_{0})=\int^{r_{0}}_{0}dr4\pi r^{2}\alpha e^{6\phi}a^{1/2}bS^{t}|_{t},
\end{equation}
where $S^{t}$ is expressed as follows:
\begin{equation}
S^{t}=E\frac{1}{\alpha}\sqrt{\frac{b}{a}}r\left(2\phi^{\prime}+\frac{b^{\prime}}{2b}+\frac{1}{r}\right)-p_{r}\frac{1}{\alpha}\sqrt{\frac{b}{a}}r\left(\frac{1}{3}K+\frac{B}{b}\right).
\end{equation}
The conservation law can be rewritten as follows:
\begin{equation}
\frac{\partial}{\partial t}\left\{M(t,r_{0})+P(t,t_{0},r_{0})\right\}=0,
\end{equation}
where $P(t,t_{0},r_{0})$ is the integrated energy flux through the sphere of the radius $r_{0}$, defined by
\begin{equation}
P(t_{1},t_{2};r_{0})=\int^{t_{2}}_{t_{1}}dt4\pi\alpha e^{6\phi}a^{1/2}br_{0}^{2}S^{r}|_{r_{0}}.
\end{equation}
$S^{r}$ is given as follows:
\begin{widetext}
\begin{eqnarray}
S^{r}&=&-E\frac{\beta}{\alpha}\sqrt{\frac{b}{a}}r\left(2\phi^{\prime}+\frac{b^{\prime}}{2b}+\frac{1}{r}\right)+E\frac{\beta^{2}}{\alpha^{2}}e^{4\phi}\sqrt{ab}r\left(\frac{K}{3}+\frac{B}{b}\right)\nonumber\\
&+&p_{r}e^{-4\phi}\frac{1}{a}\sqrt{\frac{b}{a}}r\left(2\phi^{\prime}+\frac{b^{\prime}}{2b}+\frac{1}{r}\right)-p_{r}\frac{\beta}{\alpha}\sqrt{\frac{b}{a}}r\left(\frac{K}{3}+\frac{B}{b}\right)-S_{rr}e^{-4\phi}\frac{1}{a}\sqrt{\frac{b}{a}}r\left(\frac{K}{3}+\frac{B}{b}\right)\,.
\end{eqnarray}
\end{widetext}
%
%%%%%%%%%%%%%%%%%%%%%%%%%%%%%%%%%%%%%%%%%%%%%%%%%%%%%%%%%%
\subsubsection{Lifetime}\label{Sec.Definition of lifetime}
%%%%%%%%%%%%%%%%%%%%%%%%%%%%%%%%%%%%%%%%%%%%%%%%%%%%%%%%%%
In this paper, 
setting $r_0$ as the certain value
larger than the typical radius of the scalar field profile of an oscillon,
we define the lifetime $\tau$ of the oscillon as follows: 
\begin{equation}
\frac{M(\tau;r_{0})}{M{(0;r_{0})}}=\epsilon\ll 1. 
\end{equation}
That is, well before the lifetime $\tau$, 
a large part of the total energy is confined inside the sphere of the radius $r_0$, 
and the energy is radiated away from the sphere by the lifetime $\tau$. 
Hereafter we set $r_{0}=10L$
and $\epsilon=0.01$.
We checked that our results depend only very weakly on the specific values of $r_0$ and $\epsilon$.
%%%%%%%%%%%%%%%%%%%%%%%%%%%%%%%%%%%%%%%%%%%%%%%%%%%%%%%%%%%%%%%%%%%%%%%%%%%%%%%%%%%%%%%%%%%%%%%%
\subsection{Numerical scheme and convergence check}\label{Sec.Numerical scheme and convergence}
%%%%%%%%%%%%%%%%%%%%%%%%%%%%%%%%%%%%%%%%%%%%%%%%%%%%%%%%%%%%%%%%%%%%%%%%%%%%%%%%%%%%%%%%%%%%%%%%
Our numerical code is written in C++.
We use the iterative Crank-Nicolson \cite{Teukolsky:1999rm} scheme for the integration in time,
and a 2nd order finite difference method for spatial derivatives.
%Furthermore,
%i
In order to remove unphysical high frequency numerical modes,
we add the Kreiss-Oliger dissipation terms.

In the remaining part of this subsection, 
we show a result of a test simulation and the convergence of our numerical calculation by using a massless scalar case for simplicity.
The initial data of the test simulation is the following:
\begin{eqnarray}
a(t=0,r)&=&b(t=0,r)=1,\\
\Phi(t=0,r)&=&Ae^{-r^{2}/w^{2}},
\end{eqnarray}
where $A=0.2$ and $w=0.5$. 
We show the convergence of the L2 norm of the Hamiltonian constraint violation
and the conservation of the Kodama mass inside the radius $r_{0}=2.0$ in Fig.~\ref{Fig.graph_convergence_ham_norm}.
%%%%%%%%%%%%%%%%%%%%%%%%%%%%%%%%%%%%%%%%%%%%%%%%%%%%%%%%
%%%%%%%%%%%%%%%%%%%%%%%%%%%%%%%%%%%%%%%%%%%%%%%%%
\subsection{Initial data}\label{Sec-initial data}
%%%%%%%%%%%%%%%%%%%%%%%%%%%%%%%%%%%%%%%%%%%%%%%%%
In this paper, we use momentarily static Gaussian bubble and a spatially conformally flat initial data, 
which is expressed as follows:
\begin{eqnarray}
\Phi(t=0,r)&=&-\sigma+2\sigma e^{-r^{2}/R_{0}^{2}},\\
a(t=0,r)&=&b(t=0,r)=1, 
\end{eqnarray}
where $R_{0}$ is the initial parameter which corresponds to the initial radius of the bubble.
Because of the momentary static condition, the momentum constraint equation is trivially satisfied. 
The conformal factor $\phi(t=0,r)$ is determined as the solution of the Hamiltonian constraint equation.
We solve the Hamiltonian constraint equation by using a shooting method.

%%%%%%%%%%%%%%%%%%%%%%%%%%%%%%%%%%%
\section{Results}\label{Sec-Result}
%%%%%%%%%%%%%%%%%%%%%%%%%%%%%%%%%%%
%
\begin{figure*}[ht]
\begin{tabular}{cc}
\includegraphics[scale=0.65,clip]{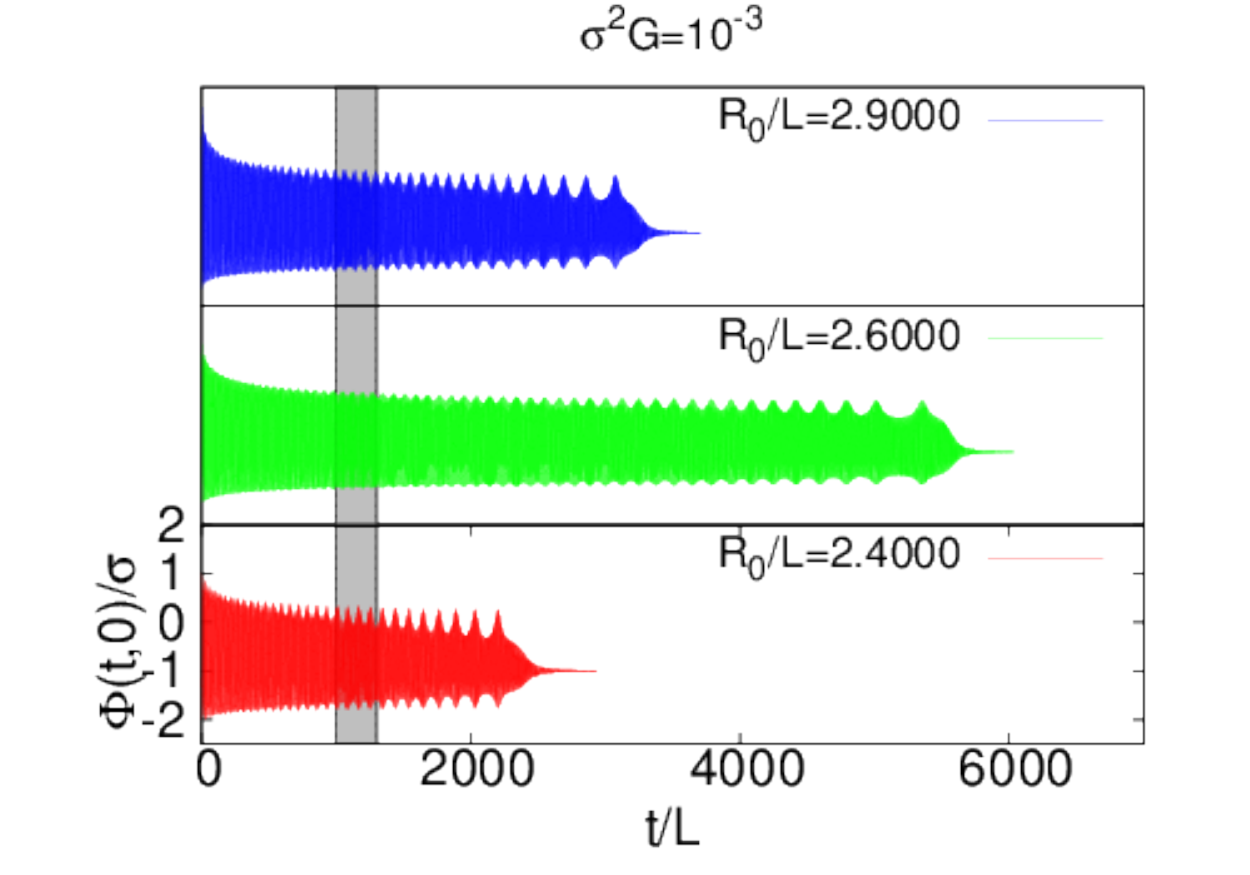}&\includegraphics[scale=0.65,clip]{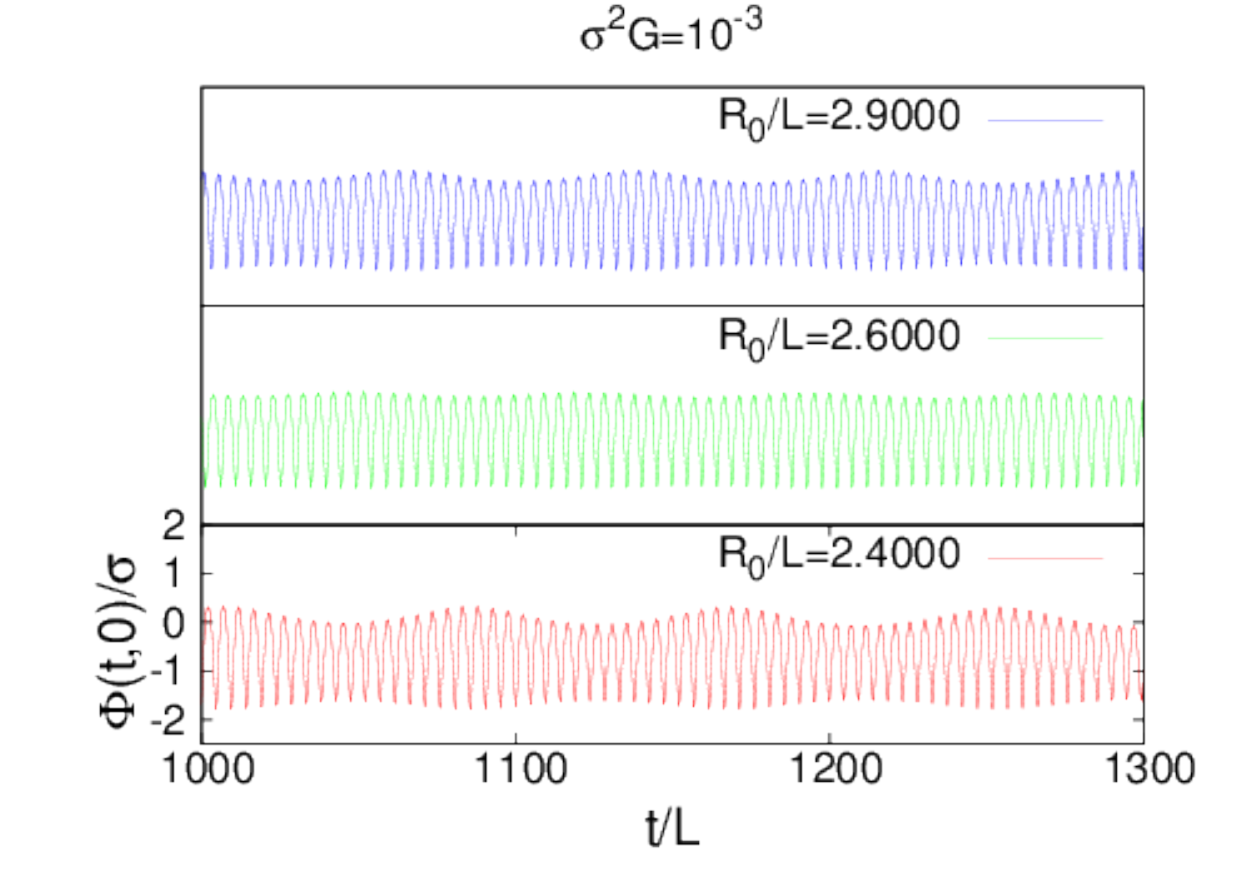}
\end{tabular}
\caption{\label{time_evolution_oscillon_0001_array}
Envelope of the time evolution of $\Phi(t,r=0)$ for $\sigma^{2}G=1.0\times 10^{-3}$. The shaded area in the left panel is zoomed in and shown in more detail on
the right panel.
}
\end{figure*}
\begin{figure}
\begin{tabular}{c}
\includegraphics[scale=0.65,clip]{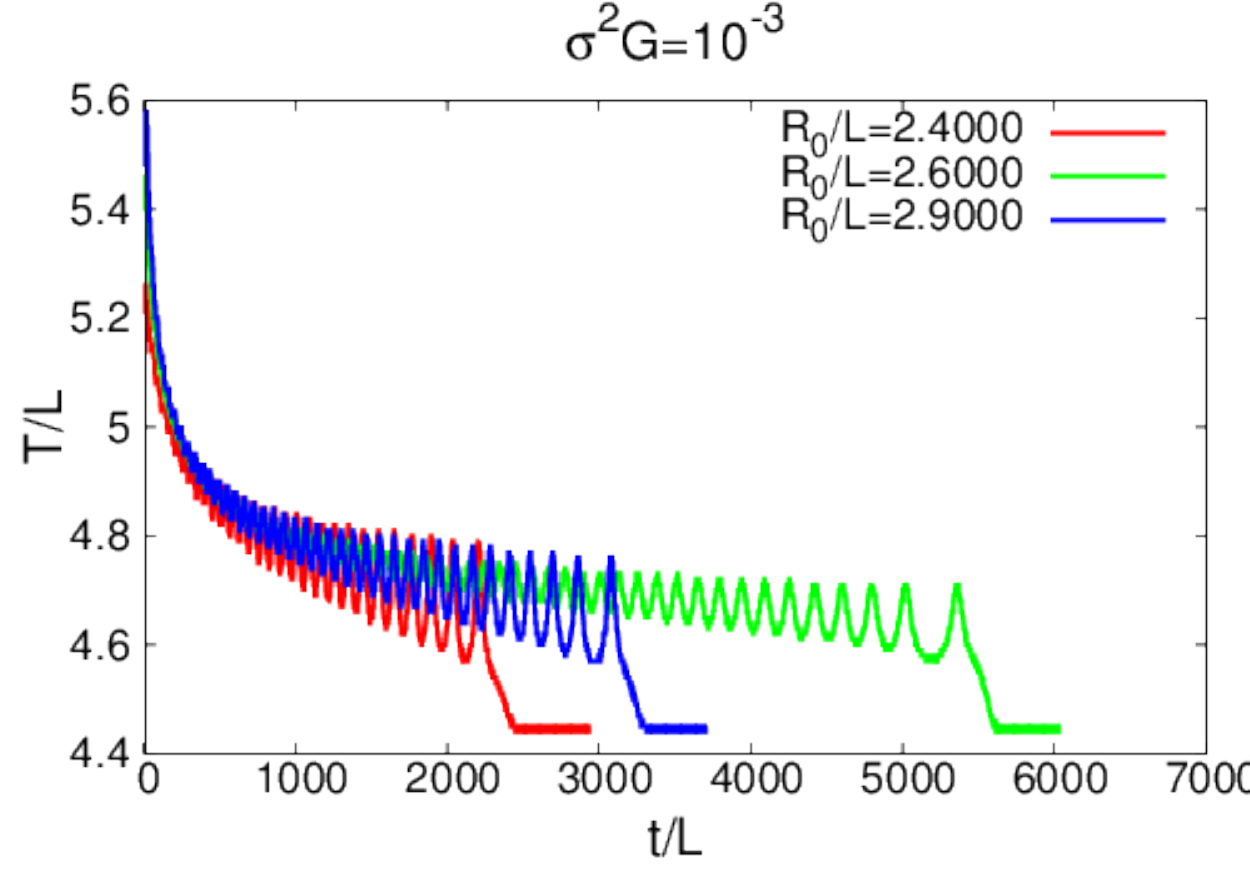}
\end{tabular}
\caption{\label{time_evolution_oscillon_period_0001}
The panel shows the time evolution of the period $T$ of the scalar field at the origin for $\sigma^{2}G=1.0\times 10^{-3}$.
}
\end{figure}
The Einstein-scalar system with the double well potential has one independent parameter $\sigma^{2}G$,
which characterizes the strength of the coupling between the scalar field and gravity.
We now study the bubble collapse, oscillon formation and its properties for $\sigma^{2}G=1.0\times 10^{-4}$, $5.0\times 10^{-4}$, $1.0\times 10^{-3}$ and $2.0\times 10^{-3}$.

We find that self-gravitating oscillons generally appear after the bubble collapses, 
and seem to have an infinite (or longer than our code can probe) lifetime at certain values of the initial bubble radius $R_0$.
Then, in Sec.~\ref{sub sec Fine structure of scaling} and~\ref{sub sec Strong gravity case}, 
we show the fine print of gravity. A new fine structure of the scaling law of the lifetime near the critical point 
is discussed in Sec.~\ref{sub sec Fine structure of scaling}, and 
the gravitational binding of the scalar field is discussed for 
the relatively strong gravity case $\sigma^{2}G=3.0\times 10^{-3}$ 
in Sec.~\ref{sub sec Strong gravity case}. 

%%%%%%%%%%%%%%%%%%%%%%%%%%%%%%%%%%%%%%%%%%%%%%%%%%%%%%%%%%%%%%%%%%%%%%%%%%%%%%%%%%
\subsection{The lifetime of oscillons}\label{sub sec Typical behavior of lifetime}
%%%%%%%%%%%%%%%%%%%%%%%%%%%%%%%%%%%%%%%%%%%%%%%%%%%%%%%%%%%%%%%%%%%%%%%%%%%%%%%%%%
A typical scalar field profile (at the origin) is shown in Fig.~\ref{time_evolution_oscillon_0001_array} for $\sigma^{2}G=1.0\times 10^{-3}$ and for different values of initial bubble radius.
The scalar profile has a high-frequency component with period $T\sim 5 L$, which is roughly dictated by the effective mass parameter $1/L$ of the scalar (as can be read off from \eqref{eq-potential}). We define the period $T$ of this high frequency mode as the time interval between two neighboring times of $\Pi=0$ and $\dot{\Pi}>0$. 
Fig.~\ref{time_evolution_oscillon_period_0001} shows the time evolution of the period $T$. 
From Fig.~\ref{time_evolution_oscillon_period_0001}, it is visible that the period $T$ of this mode decreases.
In other words, the frequency is increasing due to non-linearities and the field is able to escape the mass-generated barrier.
When $T/L$ becomes about $4.6$, the oscillon disappears and the scalar field dissipates.
%As is shown in Fig.~\ref{time_evolution_oscillon_0001_array}, 
%the scalar field at the origin oscillates with a high frequency basic oscillating mode.
%The envelop of the scalar field oscillation modulates.

Next, 
we focus on the time evolution of the Kodama mass.
\begin{figure}
\begin{tabular}{c}
\includegraphics[scale=0.65,clip]{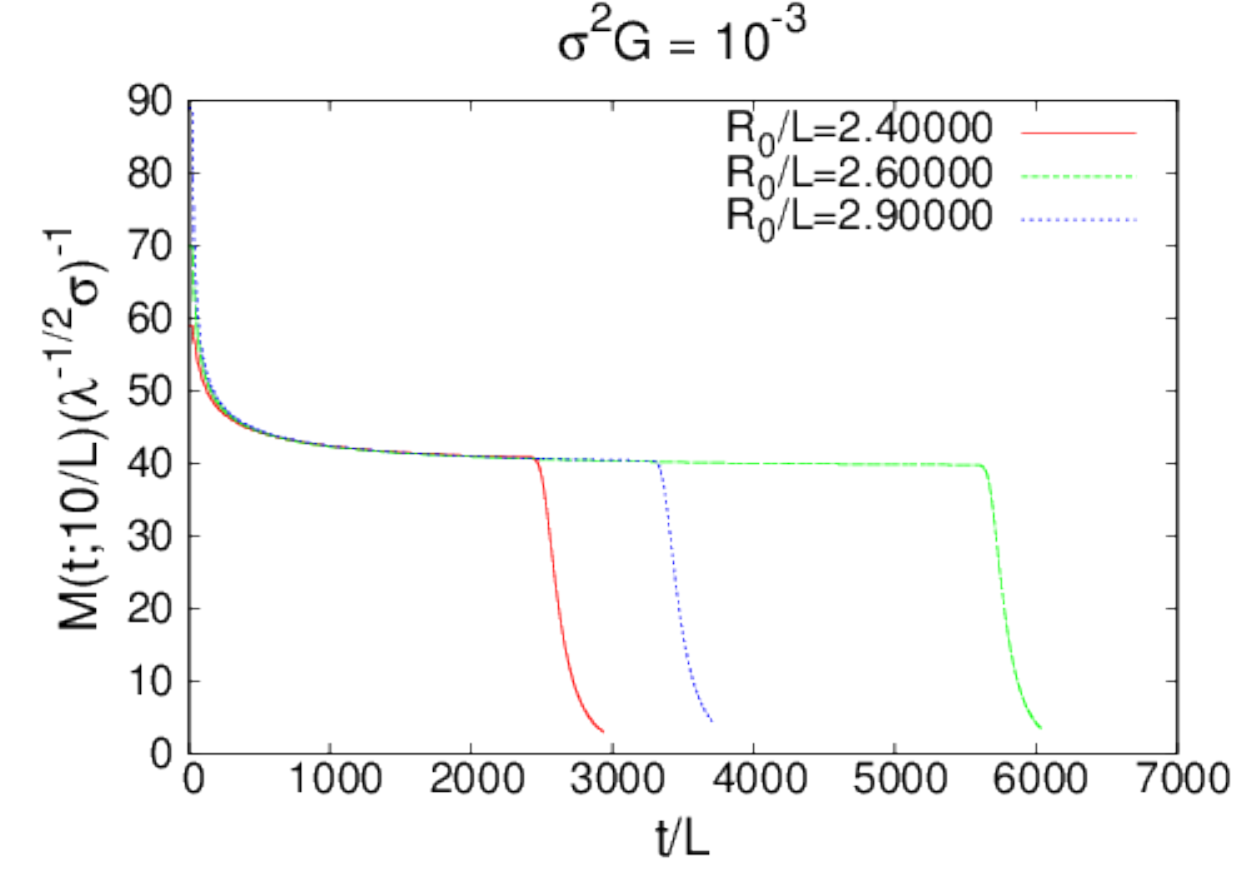}
\end{tabular}
\caption{\label{time_evolution_0001}
Time evolution of the energy inside a sphere with radius $r_{0}=10L$.
}
\end{figure}
From Fig.~\ref{time_evolution_0001}, one can see that there are mainly three stages. 
First, immediately after the collapse starts, the Kodama mass rapidly decreases due to the scalar field radiation. After this phase, the scalar field enters an ``oscillon phase'', 
and its energy (as defined by the Kodama mass inside the sphere) is almost constant, $M_{\footnotesize{\mbox{Oscillon}}}$. 
That is, the radiation of the scalar field is strongly suppressed during this stage. 

It is worth to be noted that, while $M_{\footnotesize{\mbox{Oscillon}}}$ does not depend on the initial bubble radius, 
$M_{\footnotesize{\mbox{Oscillon}}}$ depends on $\sigma^{2}G$.
The dependence is given in Table~\ref{tab:mass}.
%%%%%%%%%%%%%%%%%%%%%%%%%%%<<start table>>%%%%%%%%%%%%%%%%%%%%%%%%%%
\begin{table}[htbp]
\caption{The dependence of $M_{\footnotesize{\mbox{Oscillon}}}$ on  $\sigma^{2}G$. 
}
\label{tab:mass}
\begin{tabular}{|c|c|}
\hline
$\sigma^{2}G$&$M_{\footnotesize{\mbox{Oscillon}}}/L$\\
\hline
$1.0\times 10^{-4}$&$43$\\
\hline
$5.0\times 10^{-4}$&$41$\\
\hline
$1.0\times 10^{-3}$&$40$\\
\hline
$2.0\times 10^{-3}$&$38$\\
\hline
\end{tabular}
\end{table}
%%%%%%%%%%%%%%%%%%%%%%%%%%%<<end table>>%%%%%%%%%%%%%%%%%%%%%%%%%%
Although the reason is not clear, from the table, 
we can see that the value of $M_{\footnotesize{\mbox{Oscillon}}}$ is smaller for the larger $\sigma^{2}G$. 
After the oscillon phase, 
the scalar field dissipates to infinity. 
The lifetime of the oscillon depends on the initial bubble radius.
For $\sigma^{2}G=1.0\times 10^{-4}$ - $2.0\times 10^{-3}$,
typical lifetime of the oscillon is about $10^{3}$-$10^{4}\,L$. 
The broad features of the energy and lifetime of the oscillon 
are similar to the ones in the Minkowski background.
Furthermore, as in the case of Minkowski background,
only when the bubble radius is above a certain value and the initial energy is larger 
enough than the oscillon energy, the oscillon phase appears.

%%%%%%%%%%%%%%%%%%%%%%%%%%%%%%%%%%%%%%%%%%%%%%%%%%%%%%%%%%%%%%%%%%%%%%%%%%%%%%%%%%%%%%%%
\subsection{Fine structure of the lifetime}\label{sub sec Fine structure of life time}
%%%%%%%%%%%%%%%%%%%%%%%%%%%%%%%%%%%%%%%%%%%%%%%%%%%%%%%%%%%%%%%%%%%%%%%%%%%%%%%%%%%%%%%%
It has been shown in Ref.~\cite{Honda:2001xg} that
when the initial parameter $R_{0}$ is fine-tuned to some value $R_{\ast}$,
the lifetime of the oscillon becomes infinite. 
Close to the threshold value $R_{\ast}$,
the lifetime obeys the scaling law $\tau/L=-\gamma\ln |R_{0}-R_{\ast}|+C$, with $\gamma$ a constant independent
of whether the threshold value is approached from the left or from the right.

\begin{figure*}
\begin{tabular}{cc}
 \includegraphics[scale=0.65,clip]{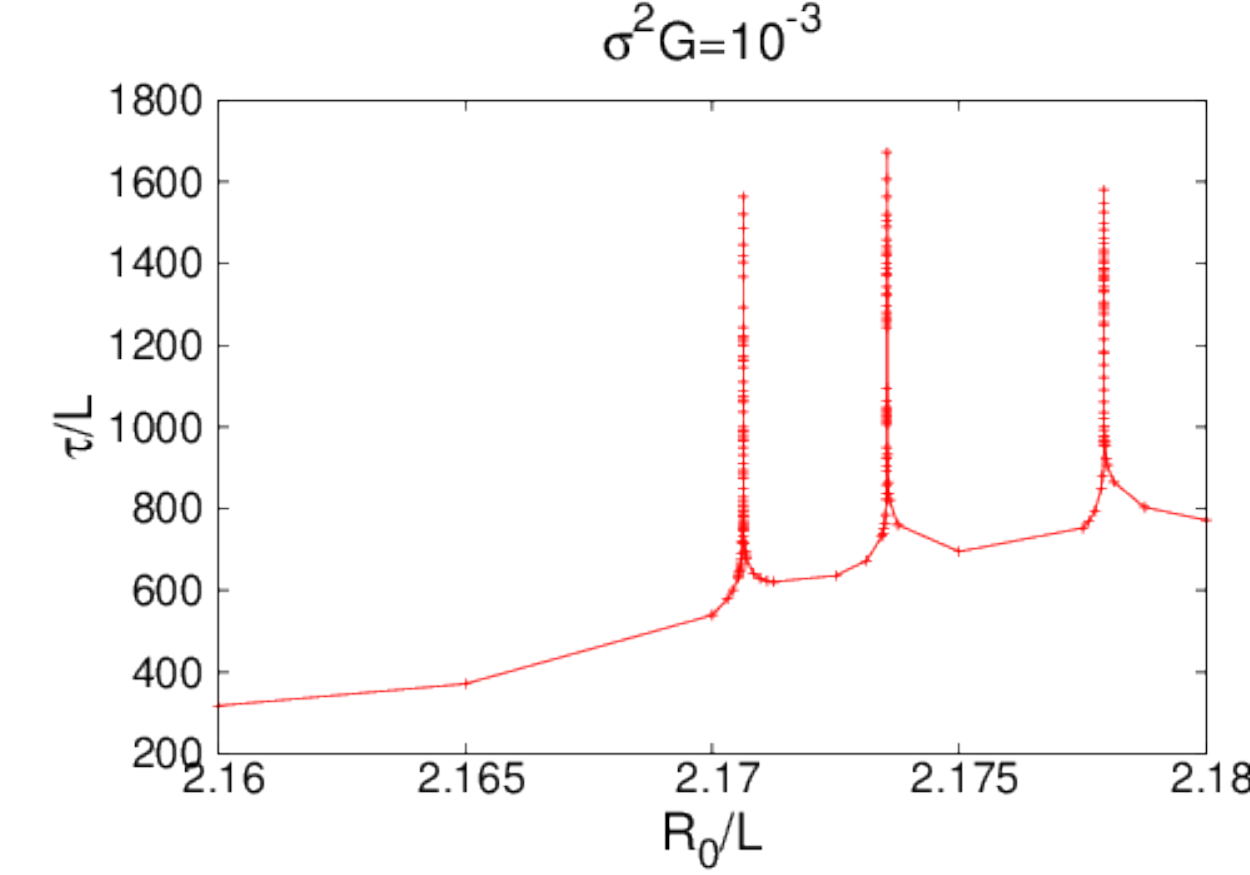}&\includegraphics[scale=0.65,clip]{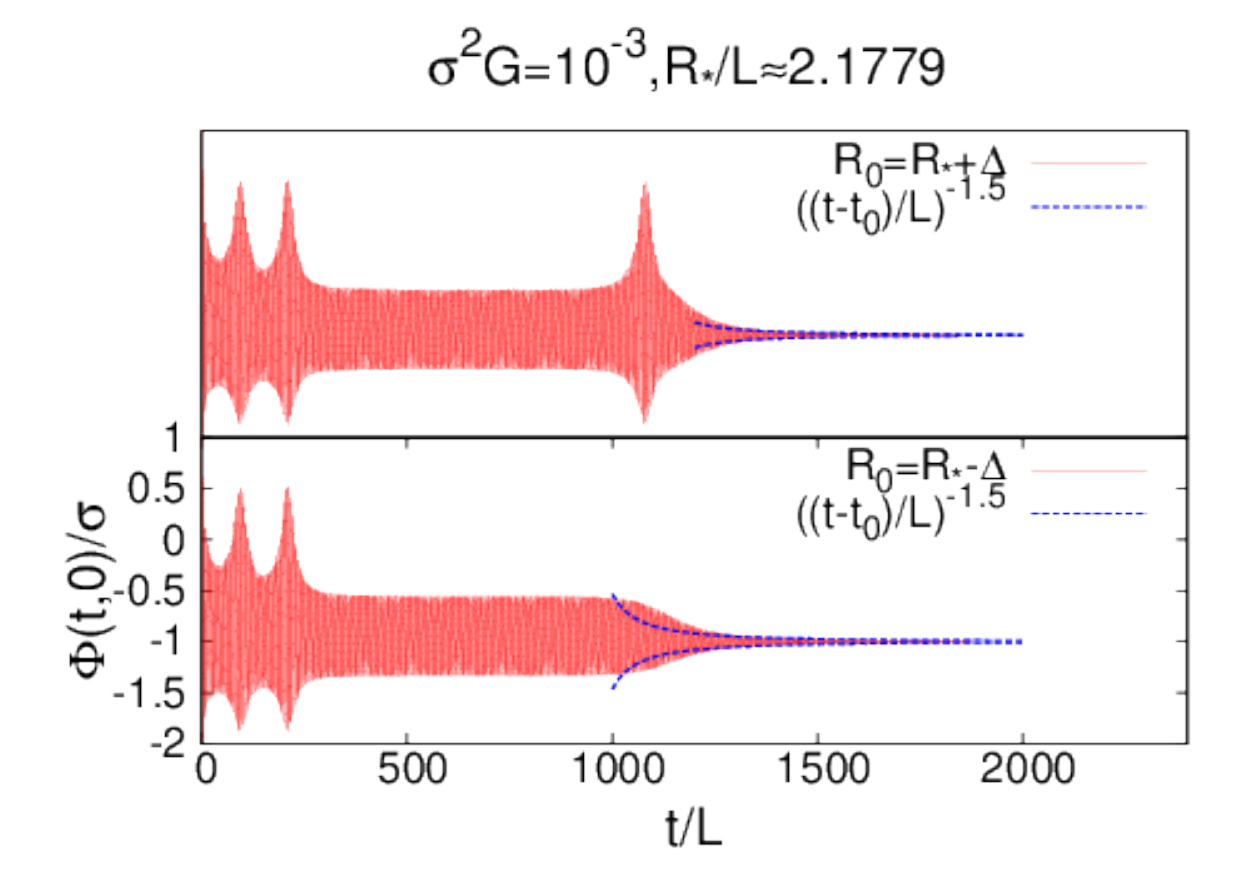}
\end{tabular}
\caption{\label{Fig.graph-time_evolution}
Evolution of oscillons for $\sigma^{2}G=10^{-3}$. 
Left panel: dependence of oscillon lifetime on initial bubble radius.
Right Panel: envelop of the time evolution of the scalar field at the origin near the third peak.
The dotted blue lines show power law dissipation of $\propto t^{-3/2}$, the expected power-law decay for massive scalars in Minkowski backgrounds.
}
\end{figure*}

We observe that the details of the evolution do depend on whether $R_{0}<R_{\ast}$ or $R_{0}>R_{\ast}$.
In particular, the number of ``modulation peaks'' seen in Fig.~\ref{time_evolution_oscillon_0001_array} depends on whether the threshold is approached from the left or from the right.
If $R_{0}$ is fine-tuned to a vicinity of $R_{\ast}$, the long-term behavior of the modulation-peaks
depends on whether $R_{0}<R_{\ast}$ or $R_{0}>R_{\ast}$.
On one side of $R_{\ast}$ the scalar field dissipates soon after this period 
while, on the other side, the envelop modulates once again just before dissipation. 
This behavior is seen clearly in Fig.~\ref{Fig.graph-time_evolution},
and is similar to the type I critical collapse (see \cite{Gundlach:2007gc}).
Our results suggest that, if we can tune $R_{0}=R_{\ast}$ exactly,
the critical solution with an infinite lifetime appears.

To investigate the fine print of these fine-tuned solutions, we focus on the first three critical solutions for each value of $\sigma^{2}G$.
The left panel of Fig.~\ref{Fig.graph-time_evolution} shows the relation between the initial bubble radius $R_{0}$ and the lifetime of oscillons.
The overall behavior of the envelop is the same as the case in the Minkowski background (see right panel in Fig.~\ref{Fig.graph-time_evolution}). 
We can also find that the lifetime near each of the first three peaks obeys the similar scaling law to the case in the Minkowski background.
\begin{figure*}
\begin{tabular}{cc}
\includegraphics[scale=0.65,clip]{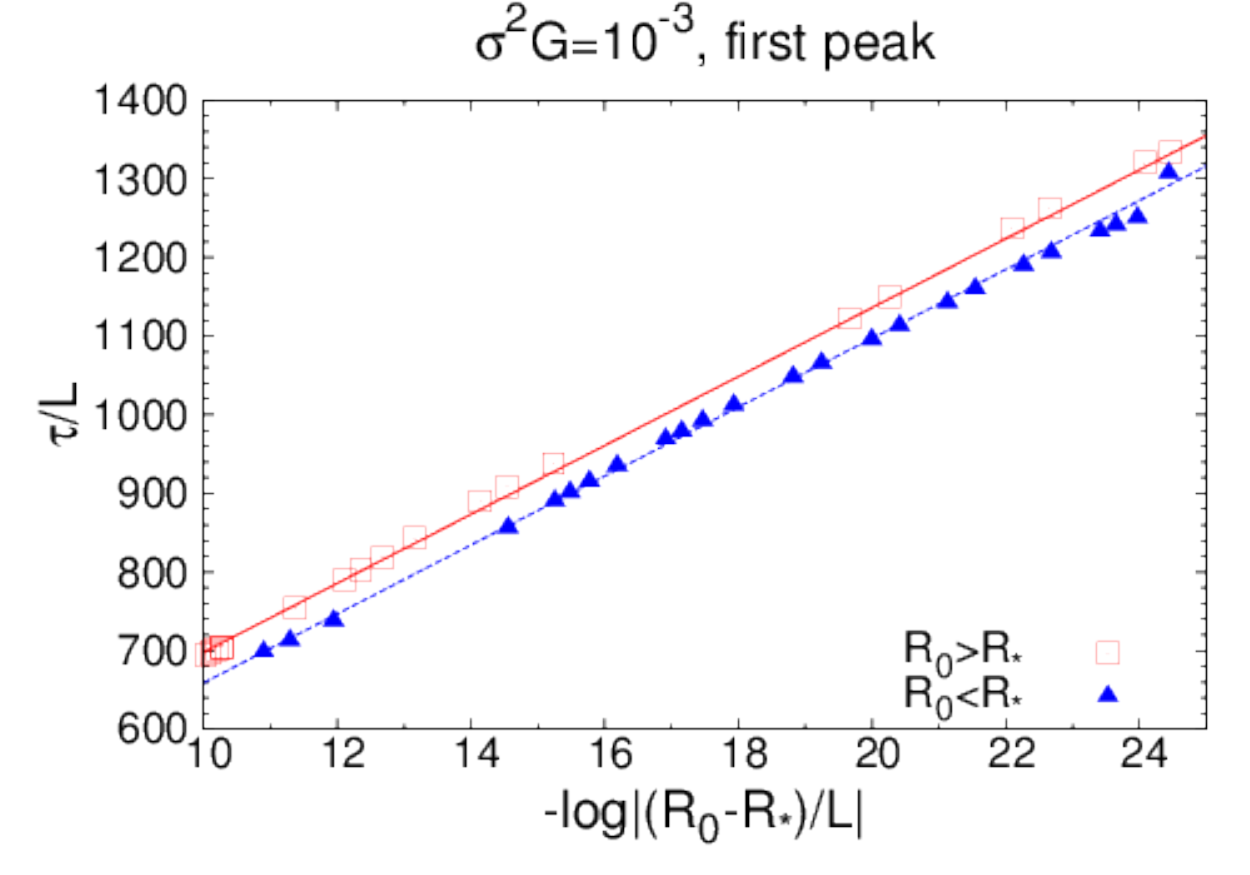}&\includegraphics[scale=0.65,clip]{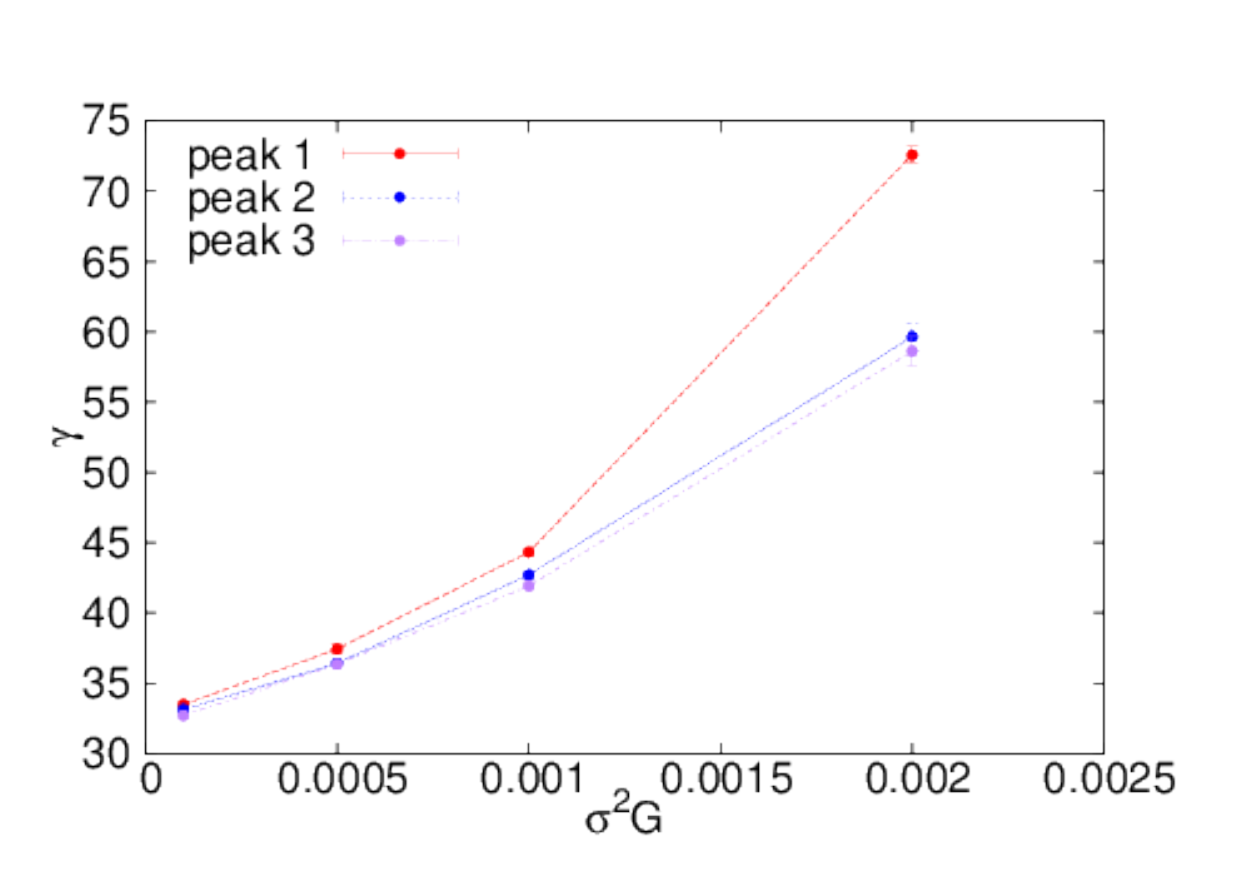}
\end{tabular}
\caption{\label{graph-relation-G-gamma}
Critical behavior in the collapse of scalars for $\sigma^{2}G=10^{-3}$. Left Panel: 
Lifetime of configuration as a function of the (log) bubble radius, close to the critical point. This study refers to the first peak in the left panel of Fig.~\ref{Fig.graph-time_evolution}. 
Notice how the broad dependence on the bubble radius is independent on whether the critical point is approached from the left or right.  
Right Panel: Relation between the exponent $\gamma$ and $\sigma^{2}G$. %\vc{The results for first peak are convergent?}
}
\end{figure*}

The index of the scaling depends on the coupling $\sigma^{2}G$ and is different for each peak.
The relation between the exponent $\gamma$ and $\sigma^{2}G$ is shown in Fig.~\ref{graph-relation-G-gamma}.

%%%%%%%%%%%%%%%%%%%%%%%%%%%%%%%%%%%%%%%%%%%%%%%%%%%%%%%%%%%%%%%%%%%%%%%%%%%%%%%%%%%%%%%
\subsection{Fine structure of the scaling law}\label{sub sec Fine structure of scaling}
%%%%%%%%%%%%%%%%%%%%%%%%%%%%%%%%%%%%%%%%%%%%%%%%%%%%%%%%%%%%%%%%%%%%%%%%%%%%%%%%%%%%%%%
%
\begin{figure*}
\begin{tabular}{cc}
\includegraphics[scale=0.65,clip]{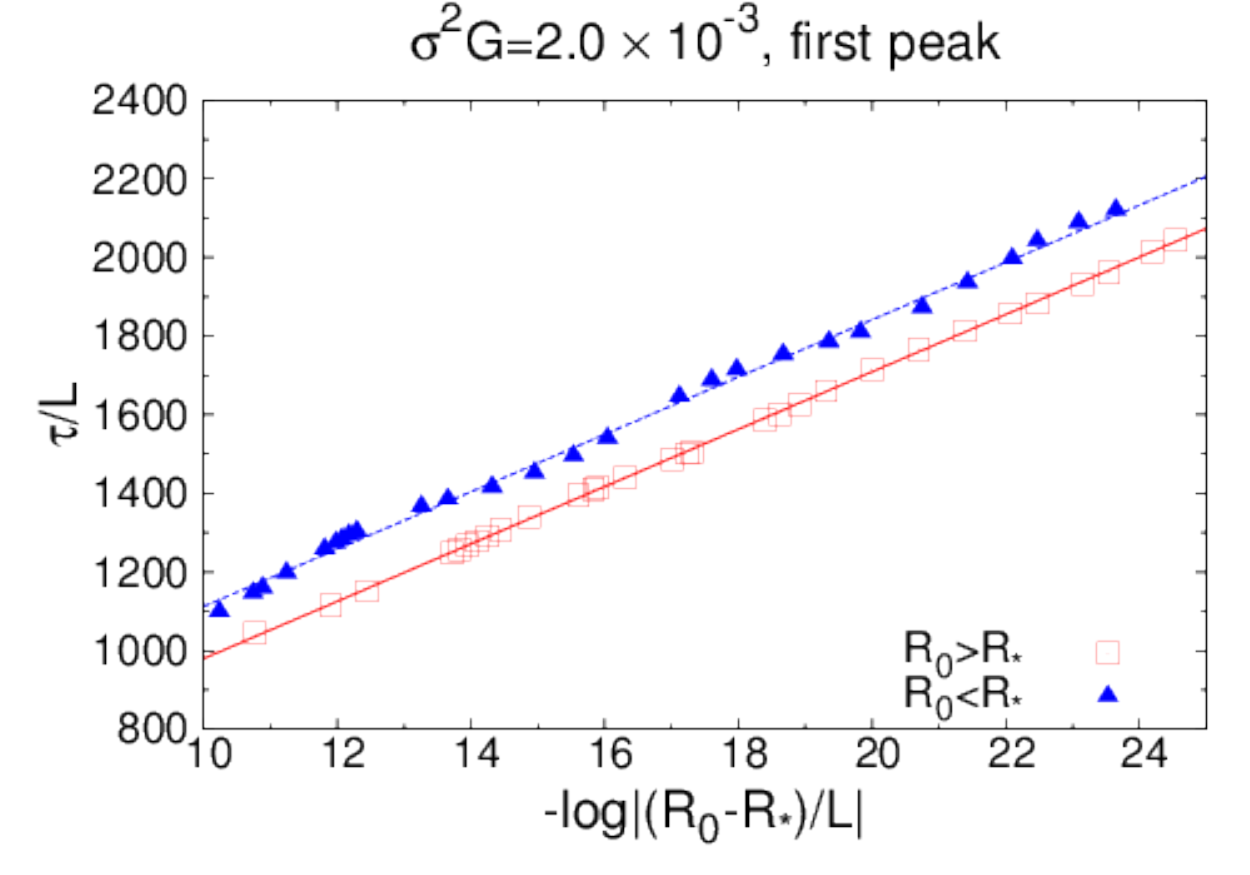}&\includegraphics[scale=0.65,clip]{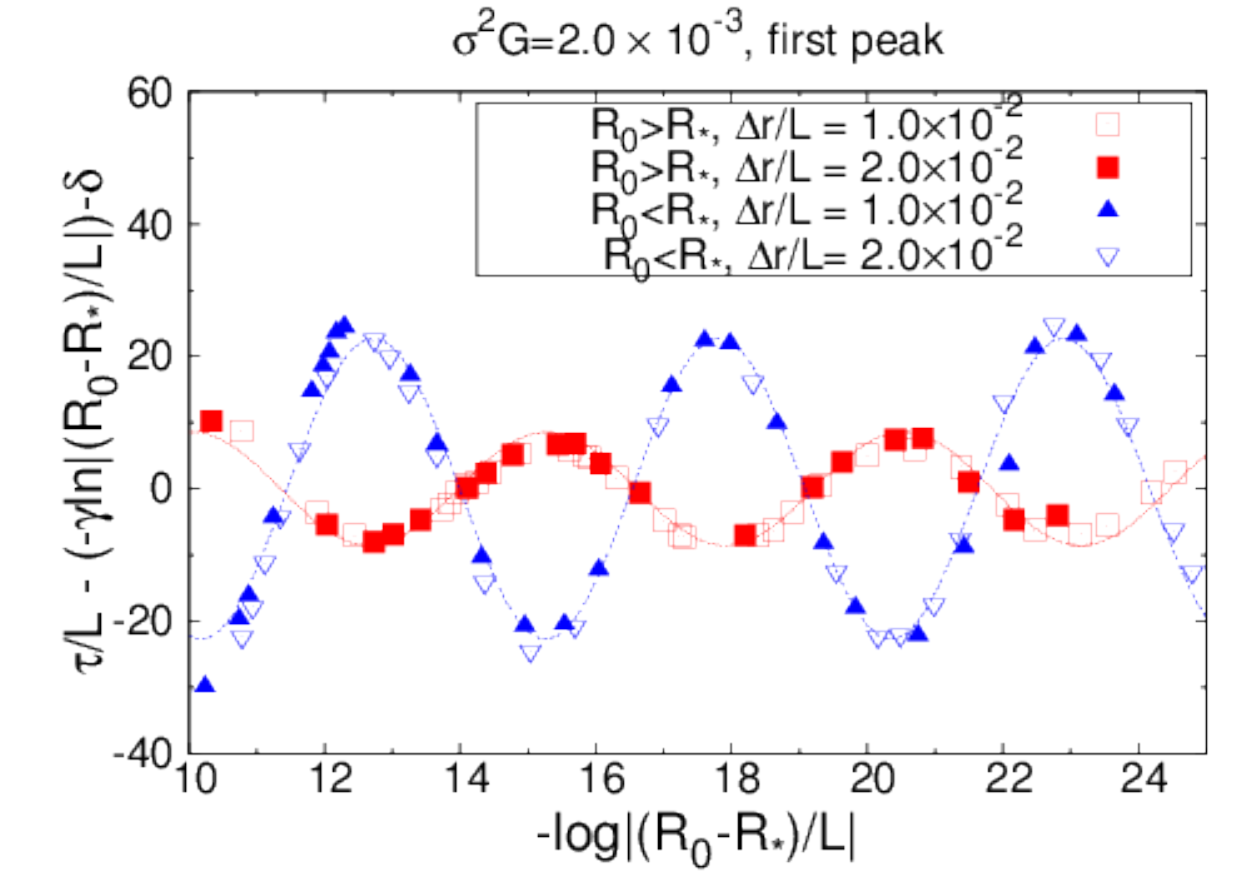}\\
\includegraphics[scale=0.65,clip]{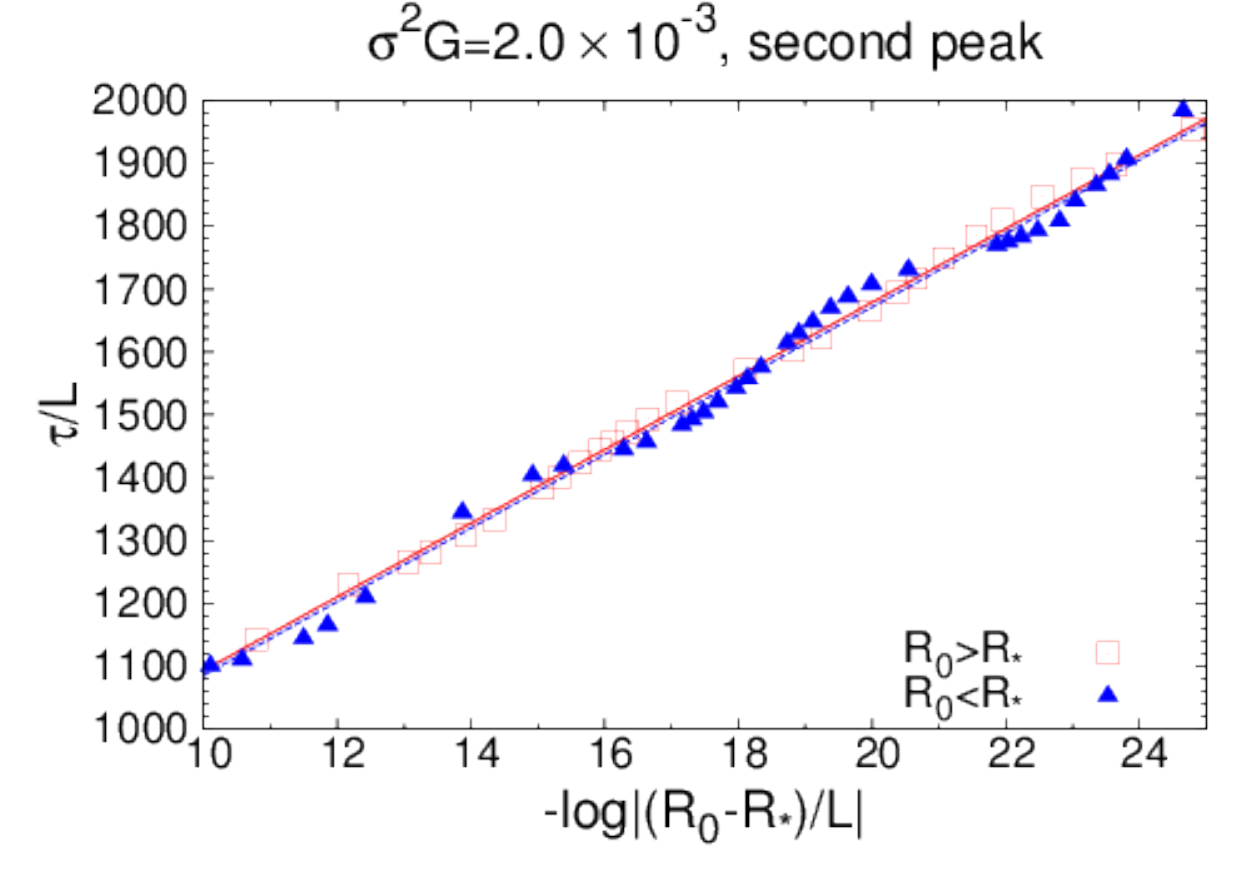}&\includegraphics[scale=0.65,clip]{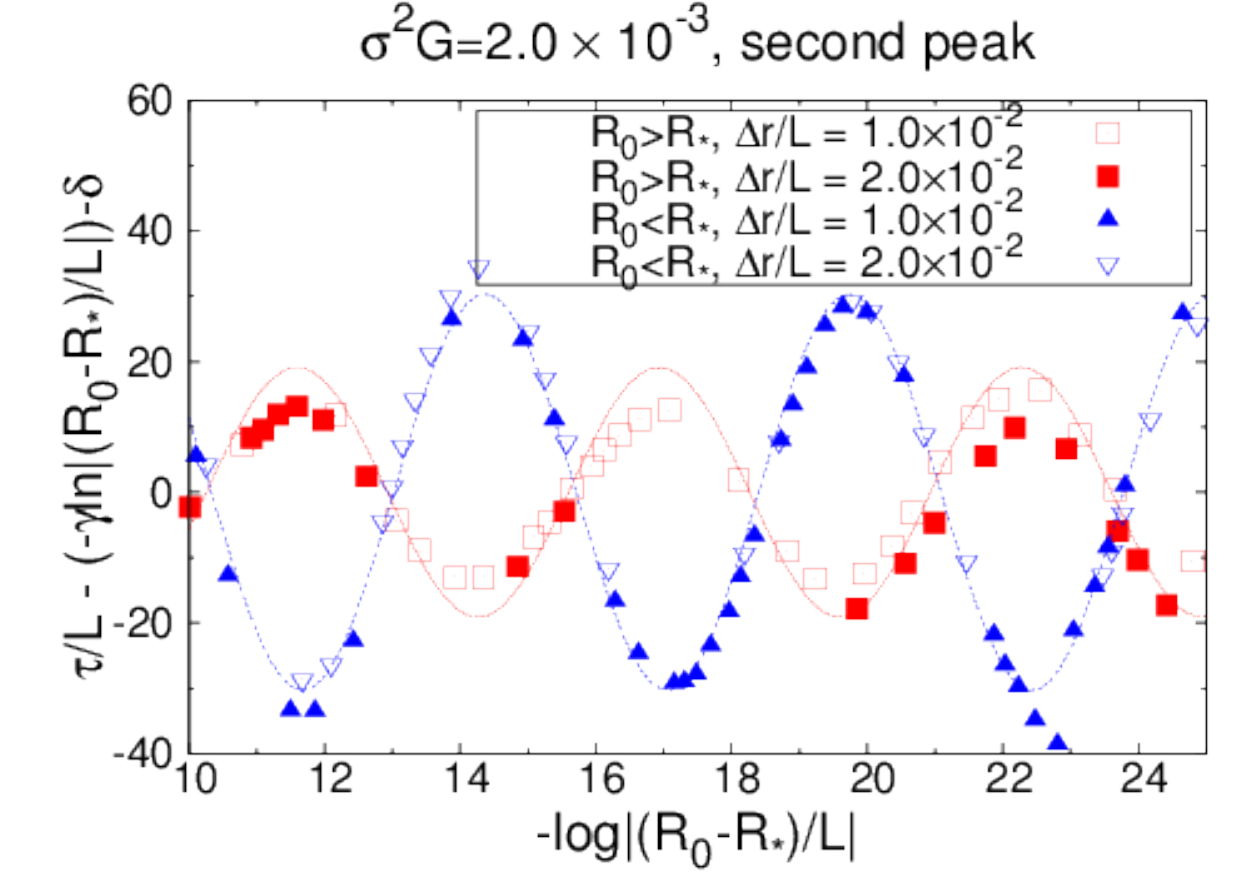}\\
\includegraphics[scale=0.65,clip]{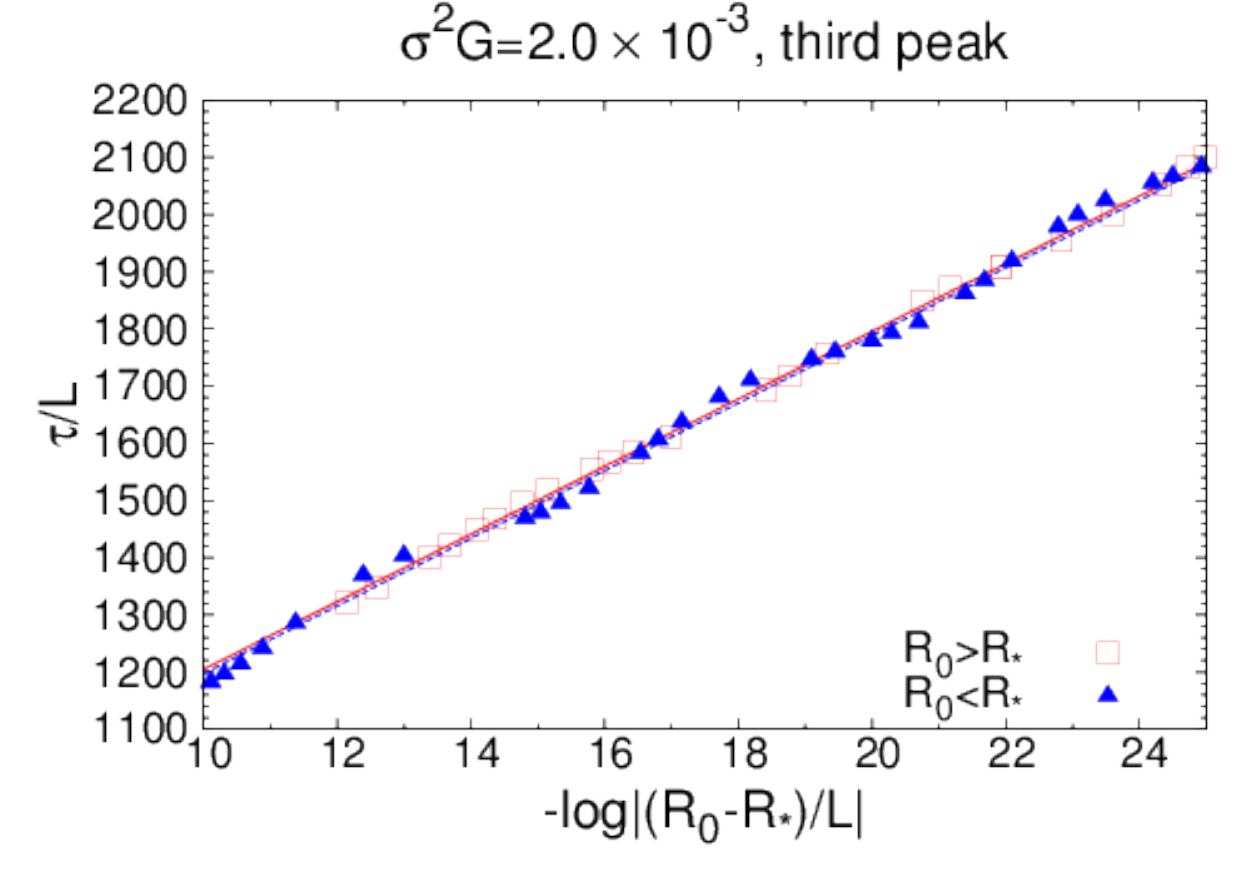}&\includegraphics[scale=0.65,clip]{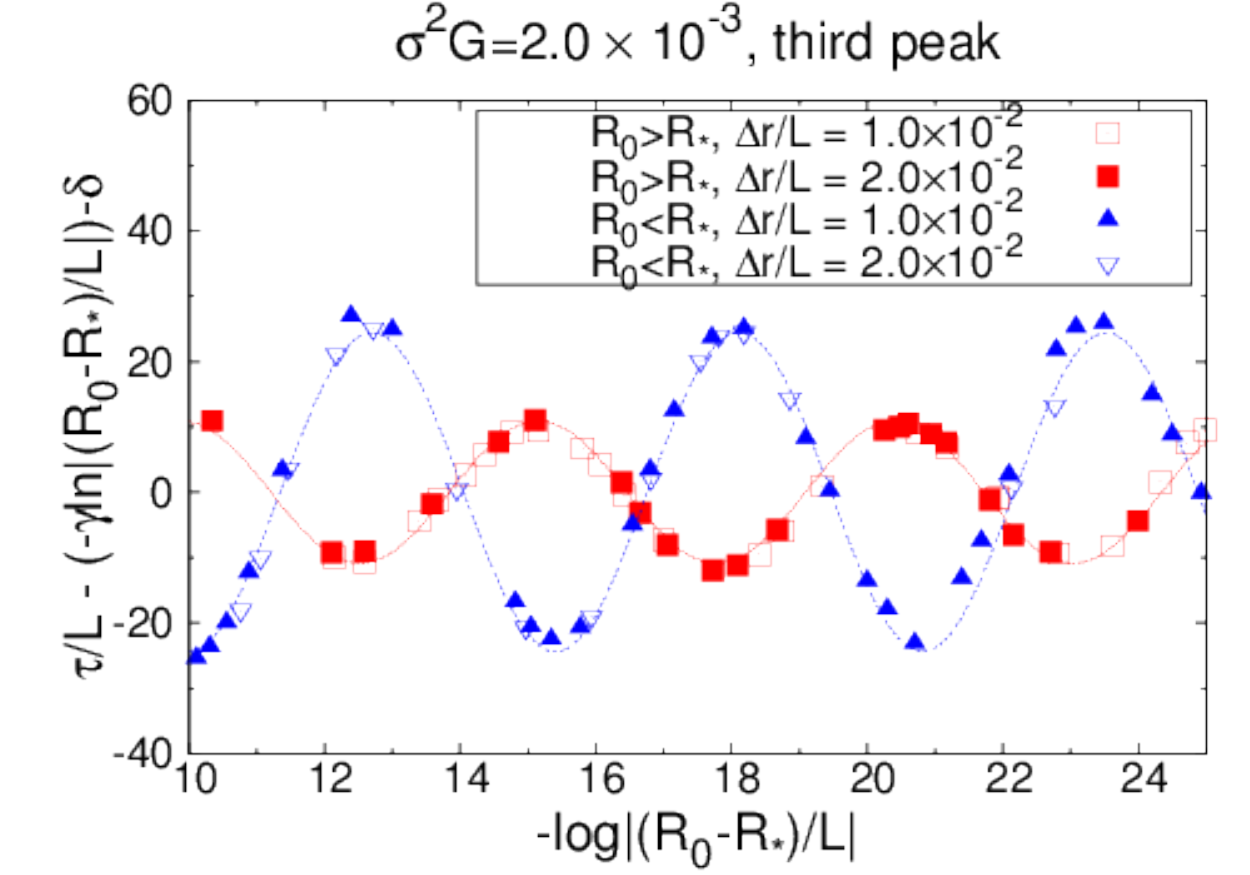}
\end{tabular}
\caption{\label{graph-Lyapunov-1-peak-0.002}
The left panels show the relation between the lifetime $\tau$ and the initial bubble radius 
near the first three peaks for $\sigma^{2}G=2.0\times 10^{-3}$. A scaling law similar to \eqref{scaling_law} is obeyed to good precision,
but there is clearly a fine structure not captured by the law.
The right panel shows the deviation between the lifetime as predicted by \eqref{scaling_law} and our actual numerical results.
To gauge the importance of grid resolution, we show results for two different grid intervals $\Delta r$. 
}
\end{figure*}
Our results show that the scaling law \eqref{scaling_law} is not the full story, and
that indeed there is a further fine structure in the dependence of the lifetime $\tau$
on the bubble radius. Our results are summarized in Figs.~\ref{graph-Lyapunov-1-peak-0.002} for $\sigma^{2} G=2.0\times 10^{-3}$.
Figures~\ref{graph-Lyapunov-1-peak-0.002} show the behavior of the lifetime near the first three peaks for $\sigma^{2} G=2.0\times 10^{-3}$ and the deviation from the simple scaling law.

The lifetime of the oscillon shows periodic modulations around the scaling law \eqref{scaling_law}, of constant amplitude. We find that the deviation
from the simple scaling law \eqref{scaling_law} is well captured by the following simple sinusoidal behavior,
\begin{eqnarray}
&&A\cos(-\log |\frac{R_{0}-R_{\ast}}{L}|+\varphi) \nonumber \\
&\equiv& \frac{\tau}{L} - (-\gamma \log |\frac{R_{0}-R_{\ast}}{L}|)-\delta\,, \label{new_scaling}
\end{eqnarray}
%
%\vc{I think there is an extra minus in the equation above...}
where $\gamma$ and $\delta$ are fixed by the fitting 
with the original scaling relation $\tau/L=-\gamma\ln |R_{0}-R_{\ast}|+C$, 
and $A$ and $\varphi$ are additional parameters, fixed by the least square fitting for 
given $\gamma$ and $\delta$. Our results indicated that this fine structure is present for other values of $\sigma^2G$, and that the amplitude $A$ depends on the magnitude of this coupling.

\begin{figure*}
\begin{tabular}{cc}
\includegraphics[scale=0.65,clip]{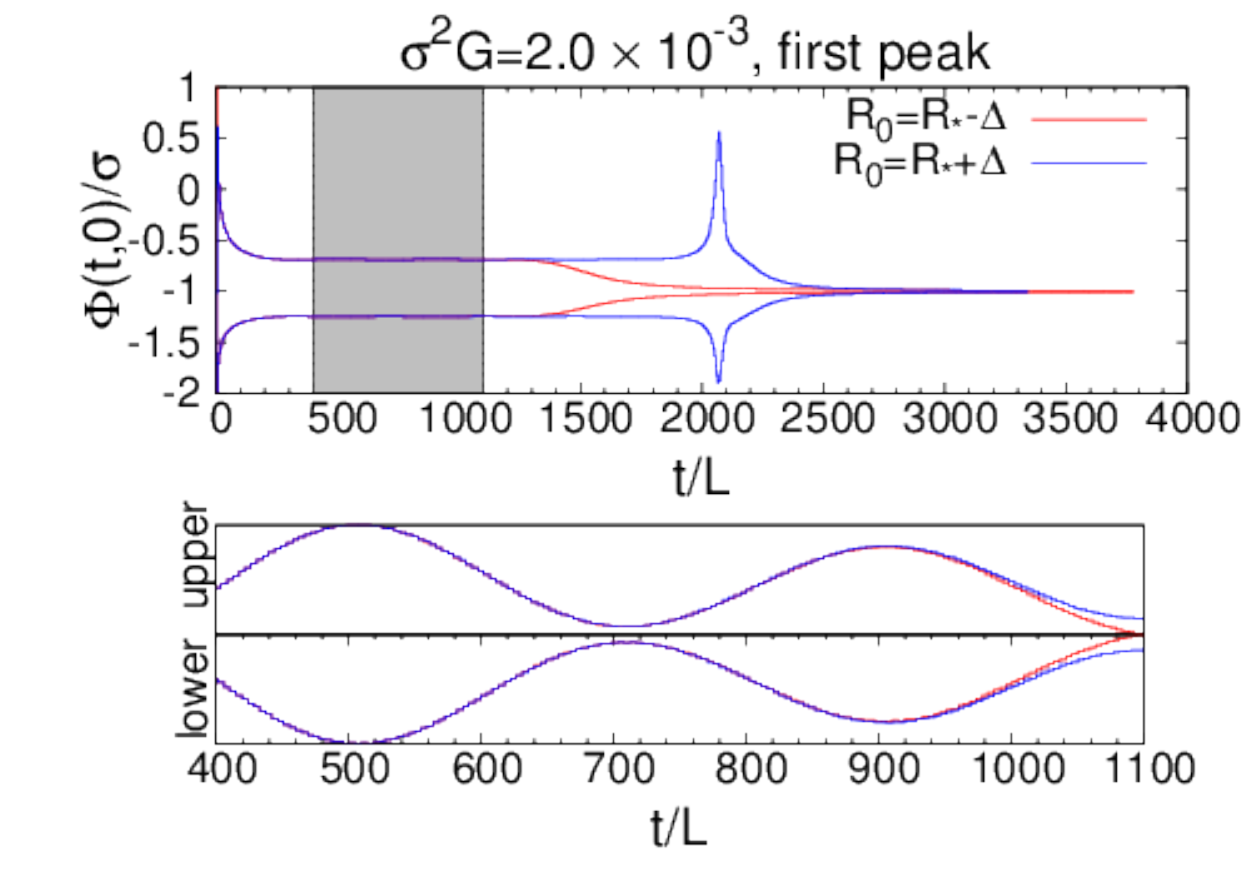}&\includegraphics[scale=0.65,clip]{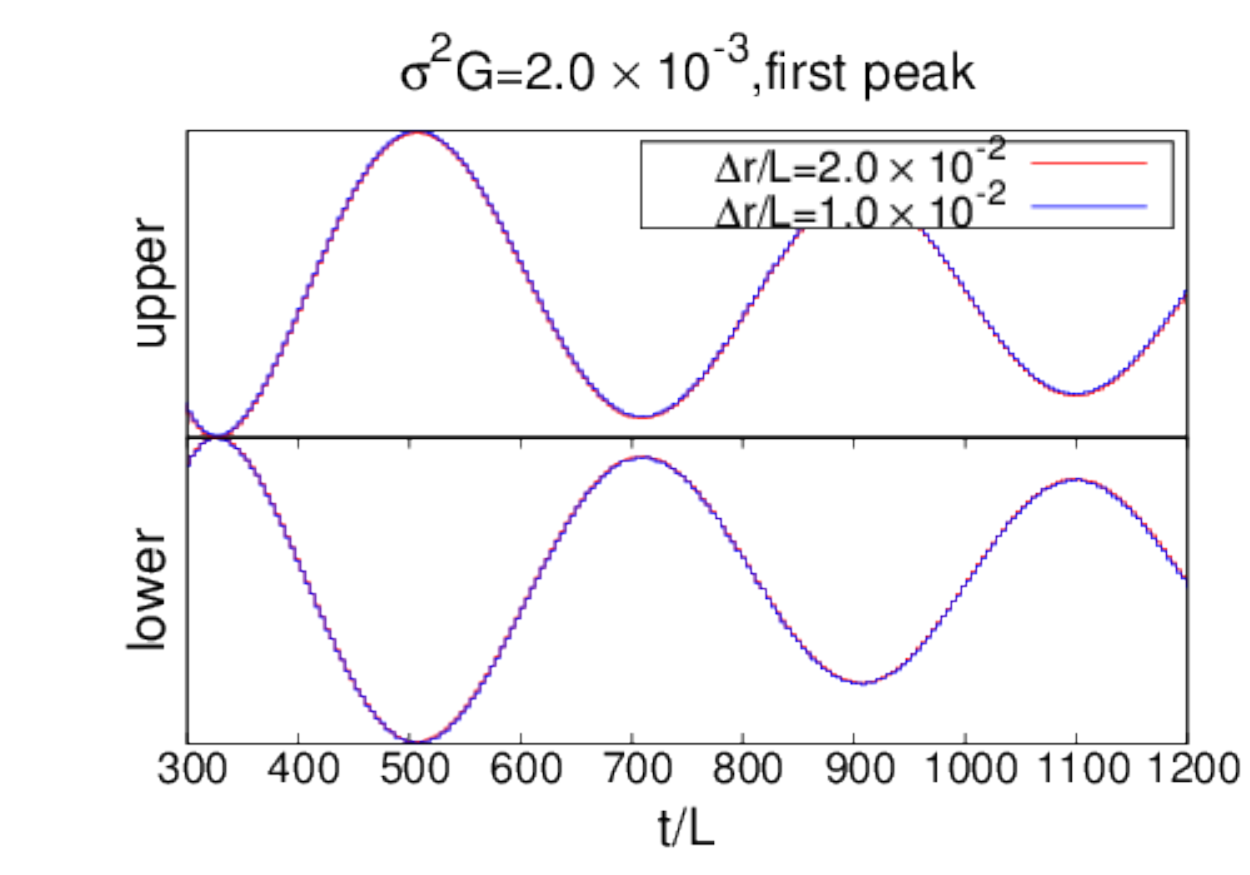}
\end{tabular}
\caption{\label{graph-time_evolution_CS_1_2}
The left upper panel shows the time evolution of the envelop of the scalar field at the origin for 
the first peak
for $\sigma^{2}G=2.0\times 10^{-3}$.
The left lower panel shows the oscillation of the envelop of the scalar field at the origin 
in the shaded region of the upper panel.
The right panel represents the convergence of the 
modulation in the plateau for different grid intervals $\Delta r$. 
}
\end{figure*}
\begin{figure*}
\begin{tabular}{cc}
\includegraphics[scale=0.65,clip]{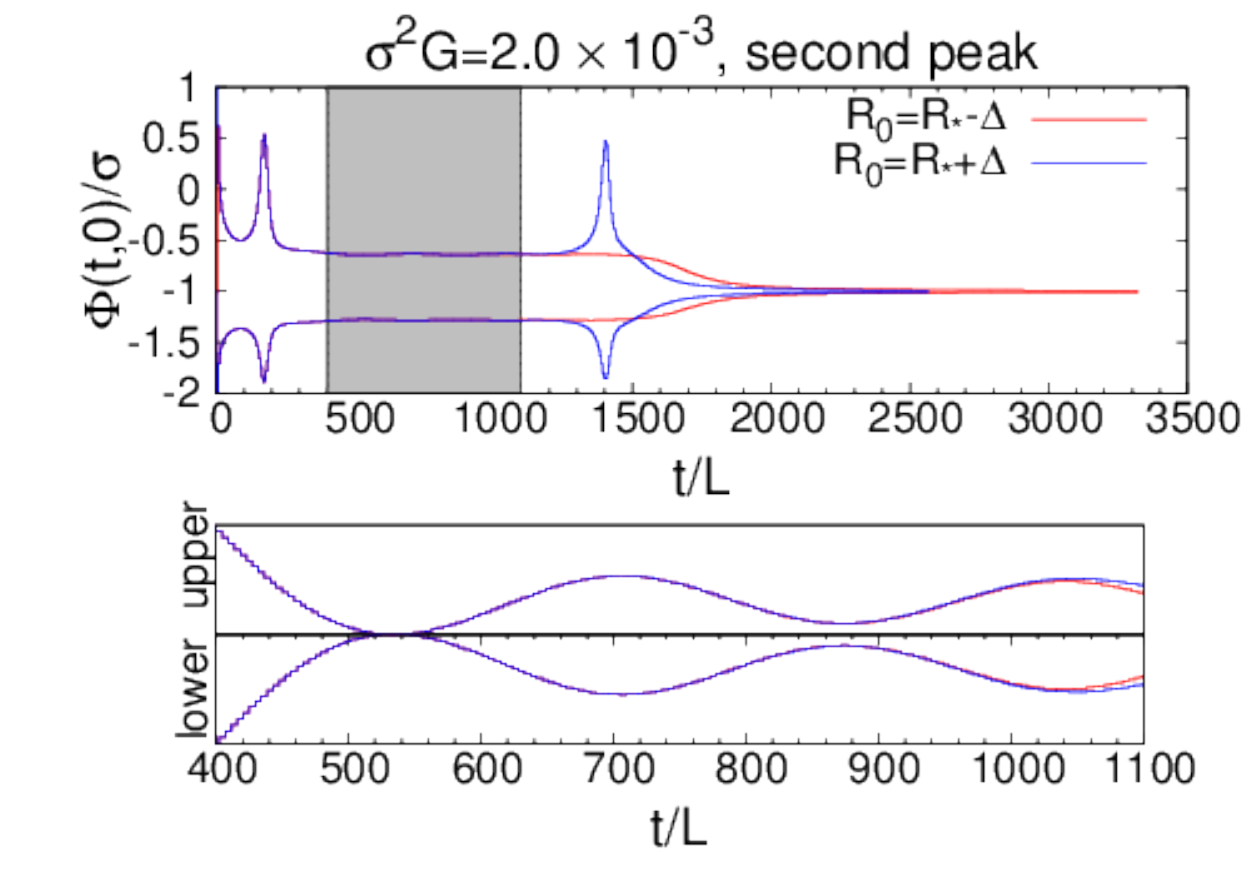}&\includegraphics[scale=0.65,clip]{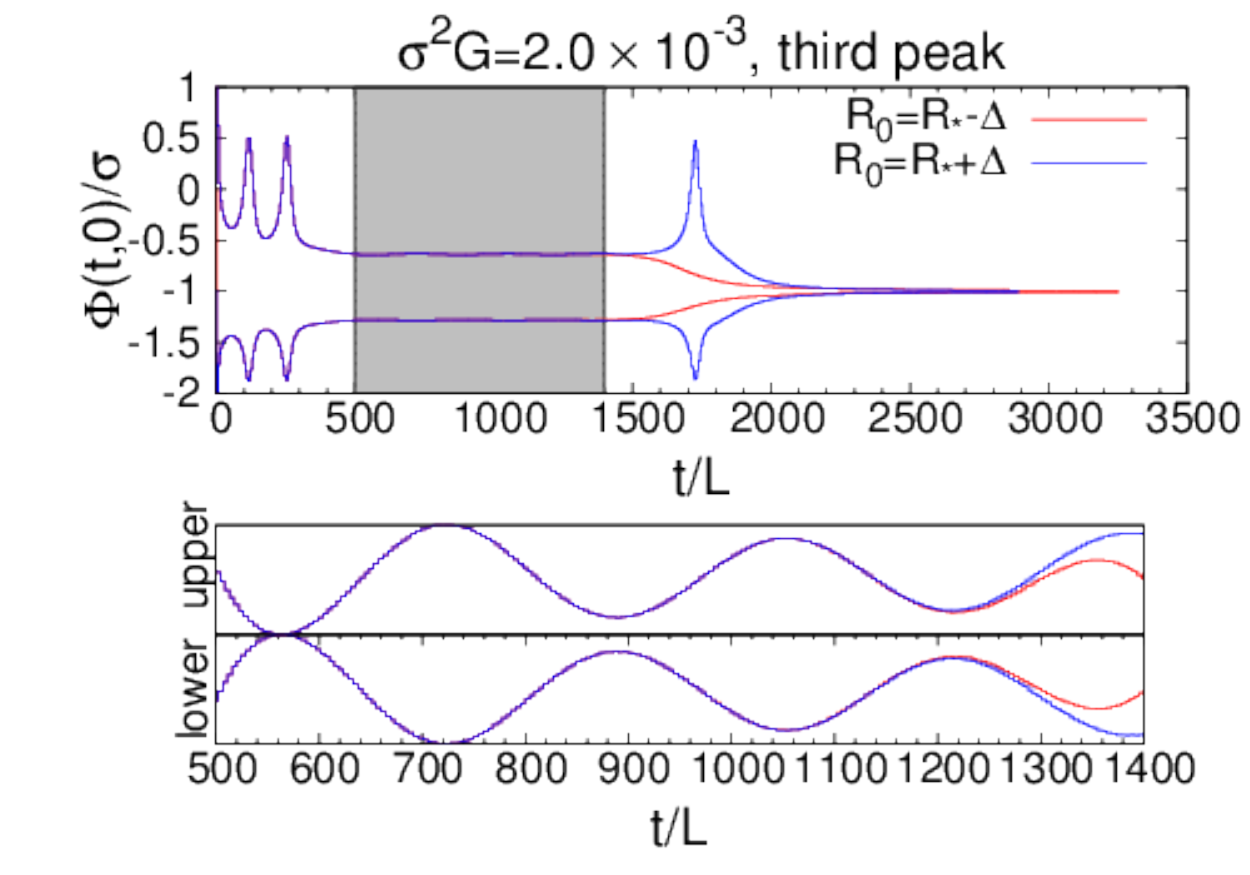}
\end{tabular}
\caption{\label{graph-time_evolution_CS_2_2}
The left upper panel shows the time evolution of the envelop of the scalar field at the origin for 
the second peak
for $\sigma^{2}G=2.0\times 10^{-3}$.
The left lower panel shows the oscillation of the envelop of the scalar field at the origin 
in the shaded region of the upper panel.
The right panel represents the convergence of the 
modulation in the plateau for different grid intervals
$\Delta r$. 
}
\end{figure*}
\begin{figure}
\begin{tabular}{c}
\includegraphics[scale=0.65,clip]{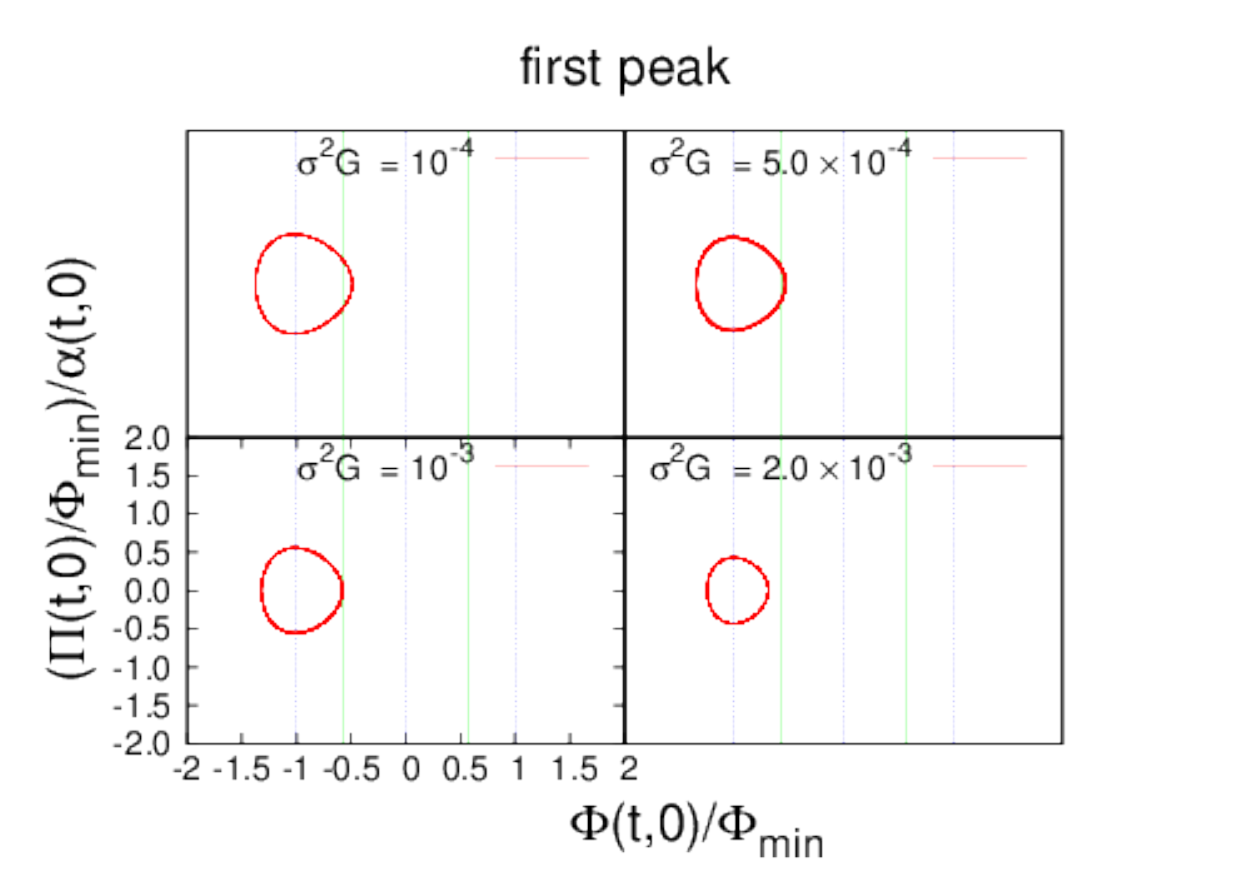}
\end{tabular}
\caption{\label{CS-section-1-peak}
The orbit of the scalar field and its conjugate momentum at the center on phase space.
The vertical blue lines and green lines correspond to $dV/d\Phi=0$ and $d^2V/d\Phi^2=0$, respectively.
}
\end{figure}

We also find a small modulation in the plateau of the envelop of the scalar field at the center, summarized in Figs.~\ref{graph-time_evolution_CS_1_2}-\ref{graph-time_evolution_CS_2_2}. 
As expected, the period of the oscillation in the fine structure of the scaling law and the period of the modulation in the plateau of the envelope are related.
From Fig.~\ref{graph-time_evolution_CS_1_2},
we can read the period $\mathcal{T}$ of the oscillation of the envelop. 
For each peak, 
period of the high frequency mode $T$, $\mathcal{T}$ and $\gamma$ are given as follows:
\begin{equation}
\begin{array}{llll}
T\simeq 5.2~~&\mathcal{T}\simeq 380~~&\gamma\simeq 73~~&\mbox{(first peak)},\\
T\simeq 5.2~~&\mathcal{T}\simeq 340~~&\gamma\simeq 59~~&\mbox{(second and third peaks)}.\\
\end{array}\nonumber
\end{equation}
The fine structure of the scaling law and the modulation of the 
plateau of the envelope are not observed for the case in the Minkowski background.
Therefore, the gravitational effect is crucial for these phenomena. 
In the period of the plateau of the envelope, 
the oscillon approximately describes the critical solution with the infinite lifetime. 
In order to see the effect of the gravity on the critical solution, 
we plot the phase space orbit of the scalar field and its conjugate momentum at the origin during the period of the plateau in Fig.~\ref{CS-section-1-peak}.
As is shown in Fig.~\ref{CS-section-1-peak}, 
we obtain smaller orbit in the phase space for the larger $\sigma^{2}G$. 
Our results suggest that the oscillation of the lifetime and the envelop of the critical solution 
are associated with a new type of type I critical behavior induced by the gravitational interaction. 

%%%%%%%%%%%%%%%%%%%%%%%%%%%%%%%%%%%%%%%%%%%%%%%%%%%%%%%%%%%%%%%%%%%%%
\subsection{Strong gravity case}\label{sub sec Strong gravity case}
\begin{figure*}
\begin{tabular}{cc}
\includegraphics[scale=0.65,clip]{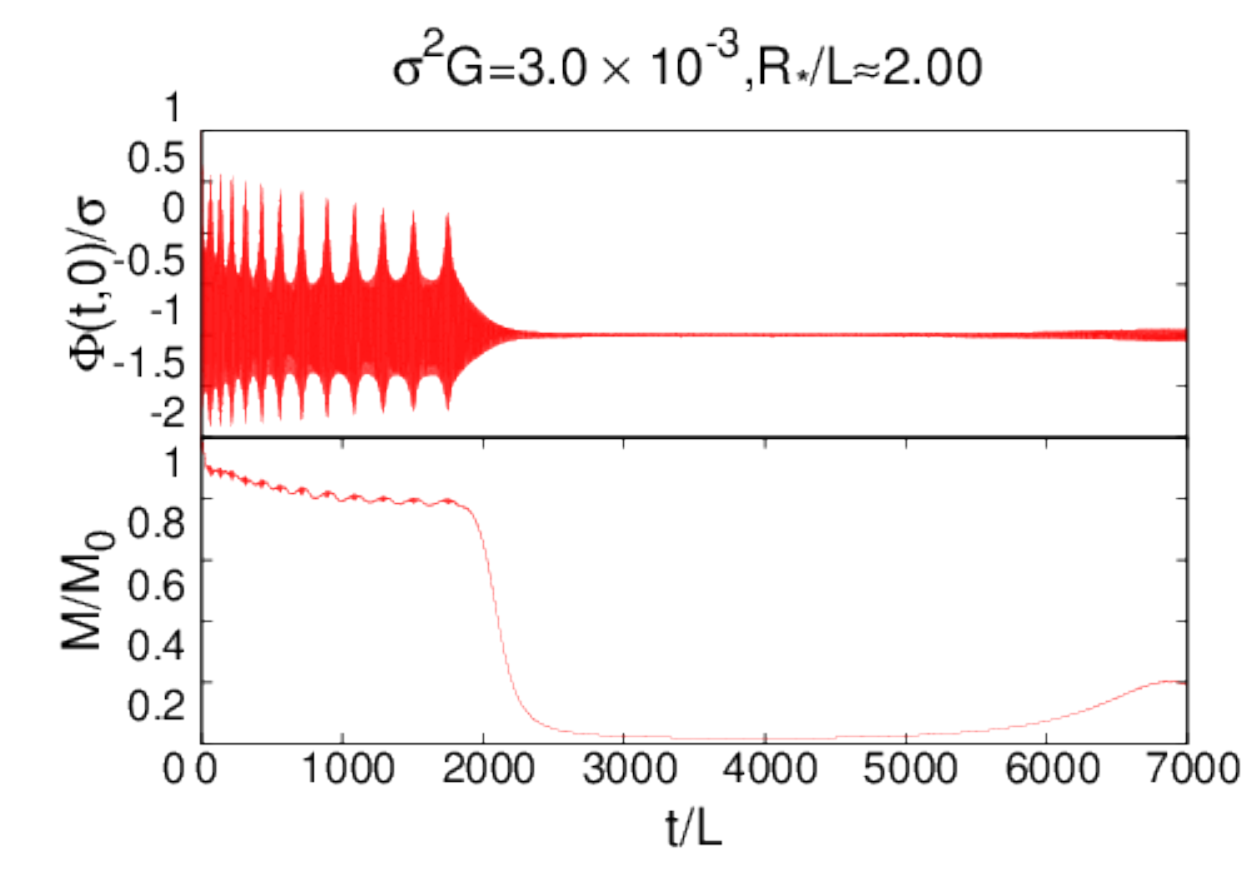}&\includegraphics[scale=0.65,clip]{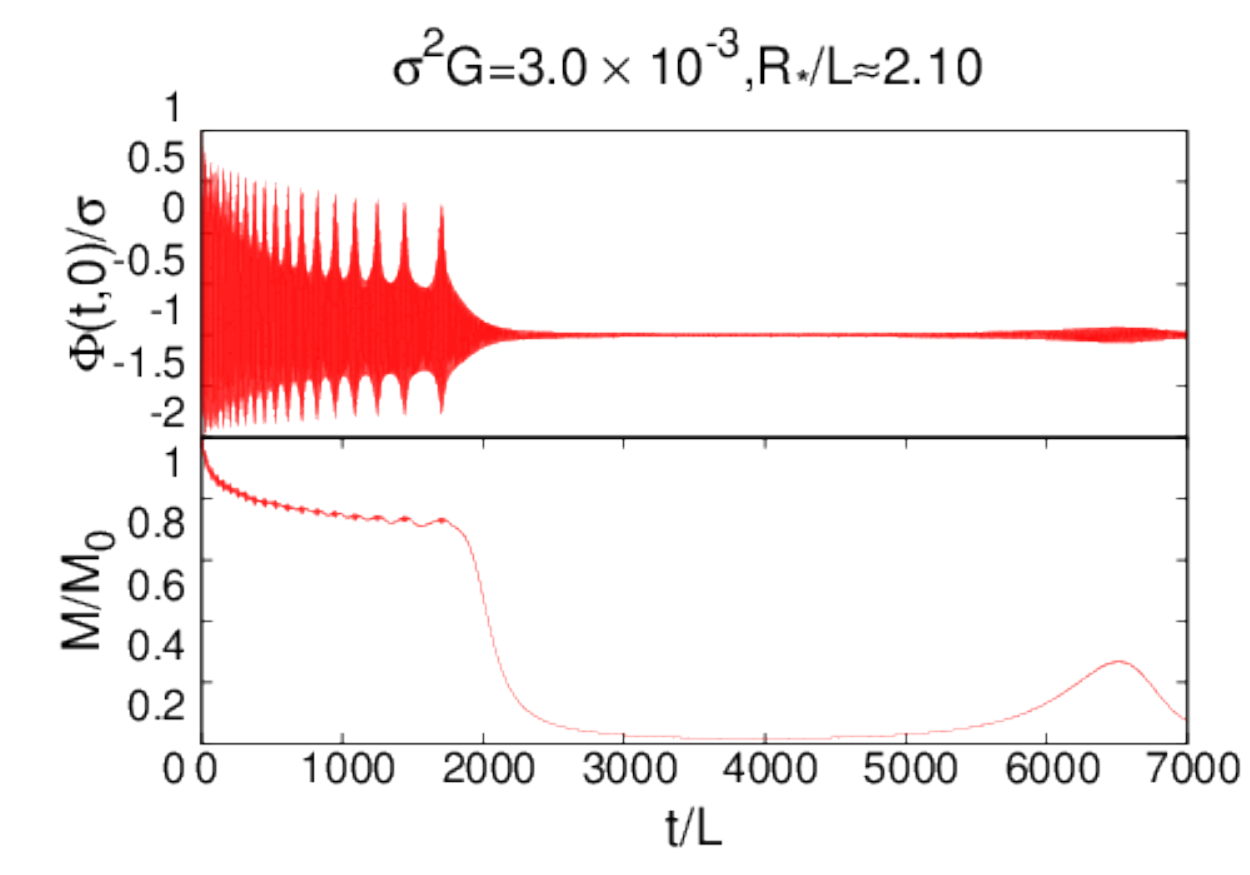}
\end{tabular}
\caption{\label{graph-time_evolution_point_strong_array_4}
The time evolution of the envelop of the
scalar field for $\sigma^{2}G=3.0\times 10^{-3}$.
The upper panel shows the envelop the scalar field at the origin.
The lower panel shows the time evolution of the Kodama mass in the sphere of the radius $r_0$.
}
\end{figure*}

%%%%%%%%%%%%%%%%%%%%%%%%%%%%%%%%%%%%%%%%%%%%%%%%%%%%%%%%%%%%%%%%%%%%%
When gravity is weak, we see no hints of anything significant happening after the power-law dispersion $\sim t^{-1.5}$ of the field
(cf. Figs.~\ref{Fig.graph-time_evolution} for example). In particular, the same late-time behavior occurs for
small couplings $10^4\sigma^{2}G=1, 5, 10, 20$, and continues at least 
until the Kodama mass is 3 orders of magnitude below its initial value. 
On the other hand, when gravity is strong we see an interesting signature. For $\sigma^{2}G>0.002$, we find cases in which 
the scalar field is gravitationally bound and eventually falls back and inside the sphere 
of radius $r_0$ at which the Kodama mass is evaluated.
Examples are shown in Fig.~\ref{graph-time_evolution_point_strong_array_4}.
Figure~\ref{graph-time_evolution_point_strong_array_4} shows this effect very clearly:
the scalar field looks to dissipate at $t\simeq 3000 L$; however, presumably due to fall-back,
it grows again at $t\simeq 5000 L$, leading to a corresponding growth of the Kodama mass.
We note that the most likely explanation for this behavior is indeed the gravitational binding energy: the scalar field value is so small around $t=1500 L$ that the non-linearity of the potential is negligible.
%%%%%%%%%%%%%%%%%%%%%%%%%%%%%%%%%%%%%%%%%%%%%%%%%%%%%%%%%%%%%%%%%%%%
\section{Summary and discussion}\label{Sec-Summary and discussion}
%%%%%%%%%%%%%%%%%%%%%%%%%%%%%%%%%%%%%%%%%%%%%%%%%%%%%%%%%%%%%%%%%%%%
We have analyzed the properties of oscillons (long-lived structures) in a spherically symmetric 
Einstein-scalar system with a double well potential. This is a very simple theory, yet with very rich phenomenology.
This system is controlled by the coupling $\sigma^{2}G$ of the scalar field to gravity.
When gravity is weak, gaussian-type initial conditions lead to collapse of the field and formation
of an oscillon, with the same properties as those observed in a Minkowski background.

When the initial bubble radius $R_{0}$ is fine-tuned to some value $R_{\ast}$, 
the lifetime $\tau$ of the oscillon is infinite. In the neighbourhood of that point, 
the lifetime obeys the scaling law $\tau/L=-\gamma\ln |R_{0}-R_{\ast}|+C$. 
We found that the index $\gamma$ depends on the coupling $\sigma^{2}G$.

Suprisingly, we also found the new characteristic features triggered by the gravitational interaction
for large couplings. The first is a new type of critical behavior, whereupon the simple scaling law above is modulated
periodically, see Eq.\eqref{new_scaling}. Likewise, the envelope of the scalar field at the origin of the critical solution oscillates. 
The periods of these oscillations are in rough agreement with each other. 
Therefore, our results suggest that the fine structure of the scaling law is reflected by the 
modulation of the oscillation of the critical solution.

In the case of a Minkowski background, even for large initial bubble radius,
the oscillon appears after collapse~\cite{Honda:2001xg}.
On the other hand, for the Einstein-scalar system, a sufficiently large initial bubble collapses to a black hole. 
Therefore, the Einstein-scalar system has richer variety of the phase space of the initial data. 
Furthermore, near the threshold of the black hole formation,
this system shows critical collapse~\cite{Ikeda:2016xmp,Clough:2016jmh}.
The critical collapse would have a variety of phases 
as in the case of the massive scalar field~\cite{Brady:1997fj, Okawa:2013jba}. 
Further investigation is needed to clarify the phase diagram of the dynamics of the Einstein-scalar system. 
We leave these for future work. 

%%%%%%%%%%%%%%%%%%%%%%%%%%%%%%%%%%%%%%%%%%%%%%%%%%%%%%%%%%%%%%%%%%%%%%%%%%%%%
\noindent{\bf{\em Acknowledgments.}}
%%%%%%%%%%%%%%%%%%%%%%%%%%%%%%%%%%%%%%%%%%%%%%%%%%%%%%%%%%%%%%%%%%%%%%%%%%%%%
We thank M. W. Choptuik for useful discussions.
This work was supported by JSPS
KAKENHI Grant Numbers JP16K17688, JP16H01097 (CY), 
KMI wakate kaigai haken program(TI) and rigaku wakate kaigai haken program(TI, CY) in Nagoya university.
V.C. acknowledges financial support provided under the European Union's H2020 ERC Consolidator Grant ``Matter and strong-field gravity: New frontiers in Einstein's theory'' grant agreement no. MaGRaTh--646597.
Research at Perimeter Institute is supported by the Government of Canada through Industry Canada and by the Province of Ontario through the Ministry of Economic Development $\&$
Innovation.
This article is based upon work from COST Action CA16104 ``GWverse'', and MP1304 ``NewCompstar'' supported by COST (European Cooperation in Science and Technology).
This work was partially supported by FCT-Portugal through the project IF/00293/2013, by the H2020-MSCA-RISE-2015 Grant No. StronGrHEP-690904.

%%%%%%%%%%%%%%%%%%%%%%%%%%%%%%%%%%%%%%%%%%%%%%%%%%%%%%%%%%
\appendix
\section{G-BSSN formulation}\label{Sec.G-BSSN formulation}
%%%%%%%%%%%%%%%%%%%%%%%%%%%%%%%%%%%%%%%%%%%%%%%%%%%%%%%%%%
In this Appendix, 
we explain the G-BSSN formulation in general coordinates.

The G-BSSN formulation is generalization of the BSSN formulation based on ADM formalism.
We assume the following metric ansatz:
\begin{equation}
ds^{2}=-\alpha^{2}dt^{2}+\gamma_{ij}(dx^{i}+\beta^{i}dt)(dx^{j}+\beta^{j}dt),
\end{equation}
where $\alpha$, $\beta^{i}$ and $\gamma_{ij}$ are the lapse function, shift vector and 3-metric, respectively.
Under this ansatz,
we can recast the Einstein equations in the evolution equations and the constraint equations:
\begin{eqnarray}
\left(\frac{\partial}{\partial t}-\mathcal{L}_{\beta}\right)\gamma_{ij}&=&-2\alpha K_{ij},\\
\left(\frac{\partial}{\partial t}-\mathcal{L}_{\beta}\right)K_{ij}&=&-D_{i}D_{j}\alpha+\alpha\{ R_{ij}+KK_{ij}-2K_{ik}K^{k}_{\ j}\nonumber\\
&+&4\pi((S-E)\gamma_{ij}-2S_{ij})\},
\end{eqnarray}
\begin{eqnarray}
R+K^{2}-K_{ij}K^{ij}&=&16\pi E,\\
D_{j}K^{j}_{\ i}-D_{i}K&=&8\pi p_{i},
\end{eqnarray}
where $K_{ij}$, $R_{ij}$, $D_{i}$ and $\mathcal{L}_{\beta}$ are the extrinsic curvature, Ricci tensor associated with $\gamma_{ij}$,  covariant derivative associated with $\gamma_{ij}$ and 
Lie derivative of $\beta^{i}$, respectively. 
%Furthermore,
$E$, $\rho_{i}$ and $S_{ij}$ are the energy density, momentum density and stress tensor of the matter sector,
which are defined as follows:
$E\equiv T_{\mu\nu}n^{\mu}n^{\nu}$, $p_{i}\equiv ^T_{\nu\mu}\gamma^{\nu}_{i}n^{\mu}$ and $S_{ij}\equiv T_{\mu\nu}\gamma^{\mu}_{i}\gamma^{\nu}_{j}$, respectively.
In the BSSN formulation,
we decompose 3-metric $\gamma_{ij}$ and the extrinsic curvature $K_{ij}$ into the following form:
\begin{eqnarray}
\gamma_{ij}&=&e^{4\phi}\tilde{\gamma}_{ij},\label{def tilde gamma}\\
K_{ij}&=&e^{4\phi}\tilde{A}_{ij}+\frac{1}{3}\gamma_{ij}K,\label{def tilde A K}
\end{eqnarray}
where $\mbox{det}(\tilde{\gamma}_{ij})=1$ and $K=\gamma^{ij}K_{ij}$.
In addition to this decomposition, 
the BSSN formulation introduces the following auxiliary field:
\begin{equation}
\tilde{\Gamma}^{k}=\tilde{\gamma}^{ij}\tilde{\Gamma}^{k}_{ij},
\end{equation}
where $\tilde{\Gamma}^{k}_{ij}$ denotes the Christoffel symbol with respect to $\tilde{\gamma}_{ij}$. 
The variables $\phi$, $\tilde{\gamma}_{ij}$, $K,~\tilde{A}_{ij}$, $\Gamma^{i}$ and 
variables of the matter sector are independent variables in the BSSN formulation. 

Although the BSSN formulation is powerful formalism in numerical relativity,
this formulation is based on Cartesian coordinates. 
In order to extend the BSSN formulation to generalized coordinates (like the spherical coordinates),
Brown introduced the G-BSSN formulation.
In the G-BSSN formulation, 
$\tilde{\gamma}\equiv\mbox{det}(\tilde{\gamma}_{ij})$ is not equal to unity. 
%
%and it is determined by the equation and its initial value.
There are two natural types of the equation to fix the value of $\tilde{\gamma}$. 
%for this equation,
One is $\partial_{t}\tilde{\gamma}=0$ (Lagrangian type),
and the other is $\partial_{\perp}\tilde{\gamma}=0$ (Eulerian type),
where $\partial_{\perp}\equiv\partial_{t}-\mathcal{L}_{\beta}$.
%Moreover,
In order to extend $\Gamma^{i}$ to the generalized coordinates,
we introduce a background metric $\bar{\gamma}_{ij}$,
and define a vector field $\Lambda^{k}$ as follows:
\begin{equation}\label{eq-def-Lambda}
\tilde{\Lambda}^{k}=\tilde{\gamma}^{ij}(\tilde{\Gamma}^{k}_{ij}-\bar{\Gamma}^{k}_{ij})=\tilde{\gamma}^{ij}\Delta\tilde{\Gamma}^{k}_{ij},
\end{equation}
where $\bar{\Gamma}^{k}_{ij}$ is the Christoffel symbol associated with $\bar{\gamma}_{ij}$.
The variables $\phi$, $\tilde{\gamma}_{ij}$, $K,~\tilde{A}_{ij}$, $\Lambda^{i}$ and variables of the matter sector
are independent variables of the G-BSSN formulation.
In this formalism, the time evolution equations can be rewritten as follows:
\begin{widetext}
\begin{eqnarray}
\left(\frac{\partial}{\partial t}-\mathcal{L}_{\beta}\right)\phi&=&-\frac{1}{6}\alpha K+\kappa\frac{1}{6}\tilde{D}_{k}\beta^{k},\label{eq-phi-evolution}\\
\left(\frac{\partial}{\partial t}-\mathcal{L}_{\beta}\right)\tilde{\gamma}_{ij}&=&-2\alpha\tilde{A}_{ij}-\kappa\frac{2}{3}\tilde{\gamma}_{ij}\tilde{D}_{k}\beta^{k},\label{eq-gamma-evolution}\\
\left(\frac{\partial}{\partial t}-\mathcal{L}_{\beta}\right)K&=&-\gamma^{ij}D_{i}D_{j}\alpha+\alpha(\tilde{A}_{ij}\tilde{A}^{ij}+\frac{1}{3}K^{2})+4\pi\alpha(E+S),\label{eq-K-evolution}\\
\left(\frac{\partial}{\partial t}-\mathcal{L}_{\beta}\right)\tilde{A}_{ij}&=&e^{-4\phi}\{-D_{i}D_{j}\alpha+\alpha(R_{ij}-8\pi S_{ij})\}^{TF}+\alpha(K\tilde{A}_{ij}-2\tilde{A}_{il}\tilde{A}^{l}_{j})-\kappa\frac{2}{3}\tilde{A}_{ij}\tilde{D}_{k}\beta^{k},\label{eq-A-evolution}\nonumber\\
\\
\left(\frac{\partial}{\partial t}-\mathcal{L}_{\beta}\right)\tilde{\Lambda}^{i}&=&\tilde{\gamma}^{mn}\bar{D}_{m}\bar{D}_{n}\beta^{i}-2\tilde{A}^{im}\partial_{m}\alpha+2\alpha(\Delta\tilde{\Gamma}^{i}_{~jk}\tilde{A}^{jk}+6\tilde{A}^{ij}\partial_{j}\phi-\frac{2}{3}\tilde{\gamma}^{ij}\partial_{j}K-8\pi\tilde{\gamma}^{ij}S_{j})\nonumber\\
&&+\frac{\kappa}{3}\{2\tilde{\Lambda}^{i}\tilde{D}_{k}\beta^{k}+\tilde{D}^{i}(\tilde{D}_{k}\beta^{k})\},\label{eq-lambda-evolution}
\end{eqnarray}
\end{widetext}
where $\mathcal{L}_{\beta}$ is the Lie derivative respect with $\beta^{i}$ and 
the superscript TF denotes the traceless part with respect to $\gamma_{ij}$.
The Ricci tensor in Eq.(\ref{eq-A-evolution}) can be expressed as follows:
\begin{eqnarray}
R_{ij}&=&R^{\phi}_{ij}+\tilde{R}_{ij},\\
\tilde{R}_{ij}&=&\tilde{\gamma}^{lm}(\Delta\tilde{\Gamma}^{k}_{~li}\Delta\tilde{\Gamma}_{jkm}+\Delta\tilde{\Gamma}^{k}_{~lj}\Delta\tilde{\Gamma}_{ikm}+\Delta\tilde{\Gamma}^{k}_{~im}\Delta\tilde{\Gamma}_{klj})\nonumber\\
&-&\frac{1}{2}\tilde{\gamma}^{lm}\bar{D}_{m}\bar{D}_{l}\tilde{\gamma}_{ij}+\frac{1}{2}(\tilde{\gamma}_{ki}\bar{D}_{j}\tilde{\Lambda}^{k}+\tilde{\gamma}_{kj}\bar{D}_{i}\tilde{\Lambda}^{k})\nonumber\\
&+&\frac{1}{2}(\tilde{\Lambda}^{k}\Delta\tilde{\Gamma}_{ijk}+\tilde{\Lambda}^{k}\Delta\tilde{\Gamma}_{jik})\,,\\
R^{\phi}_{ij}&=&-2\tilde{D}_{i}\tilde{D}_{j}\phi-2\tilde{\gamma}_{ij}\tilde{D}^{k}\tilde{D}_{k}\phi+4\tilde{D}_{i}\phi\tilde{D}_{j}\phi-4\tilde{\gamma}_{ij}\tilde{D}^{k}\phi\tilde{D}_{k}\phi.\nonumber\\
\end{eqnarray}
The value of parameter $\kappa$ depends on the choice of the time evolution of $\mbox{det}(\tilde{\gamma}_{ij})$,
$\kappa=1$ for the Lagragian option, 
and $\kappa=1$ for the Lorentzian option. 
The Hamiltonian constraint and momentum constraints are expressed as follows:
\begin{eqnarray}
&&\tilde{\gamma}^{ij}\tilde{D}_{i}\tilde{D}_{j}e^{\phi}-\frac{e^{\phi}}{8}\tilde{R}+\frac{e^{5\phi}}{8}\tilde{A}_{ij}\tilde{A}^{ij}-\frac{e^{5\phi}}{12}K^{2}+2\pi e^{5\phi}E=0,\nonumber\\
\label{eq-Ham}\\
&&\tilde{D}_{j}(e^{6\phi}\tilde{A}^{j}_{~i})-\frac{2}{3}e^{6\phi}\tilde{D}_{i}K-8\pi e^{6\phi}p_{i}=0.\label{eq-Mom}
\end{eqnarray}
%
%Because we calculate the time evolution by a free evolution scheme(see Sec.\ref{Sec.Numerical scheme and convergence}),%
The definition of $\tilde{\Lambda}^{i}$ Eq.(\ref{eq-def-Lambda}) can be regarded as an additional constraint.
These equations Eqs.(\ref{eq-phi-evolution})-(\ref{eq-lambda-evolution}) and Eqs.(\ref{eq-Ham})-(\ref{eq-Mom}) and equations of motion for the matter sector are all the equations of this system. 
%%%%%%%%%%%%%%%%%%%%%%%%%%%%%%%%%%%%%%%%%%%%%%%%%%%%%%%%%%%%%%%%%%%%%%%%%%%%%%%%%%%%%%%%%%%%%%%%%%%%%%%%%%%%%%%%%%%%%%%
\section{Explicit expressions for $\mathcal{B},~\mathcal{D}_{rr},~\mathcal{D}_{\theta\theta},~R_{rr},R_{\theta\theta}$}
\label{expressions}
%%%%%%%%%%%%%%%%%%%%%%%%%%%%%%%%%%%%%%%%%%%%%%%%%%%%%%%%%%%%%%%%%%%%%%%%%%%%%%%%%%%%%%%%%%%%%%%%%%%%%%%%%%%%%%%%%%%%%%%
$\mathcal{B},~\mathcal{D}_{rr},~\mathcal{D}_{\theta\theta},~R_{rr},R_{\theta\theta}$ in the
evolution equations are respectively given as follows: 
\begin{eqnarray}
\mathcal{B}&=&\tilde{D}_{k}\beta^{k}=\beta^{\prime}+\left(\frac{a^{\prime}}{2a}+\frac{b^{\prime}}{b}+\frac{2}{r}\right)\beta,\\
\mathcal{D}_{rr}&=&\alpha^{\prime\prime}-\left(\frac{a^{\prime}}{2a}+2\phi^{\prime}\right)\alpha^{\prime},\\
\mathcal{D}_{\theta\theta}&=&r\alpha^{\prime}\frac{b}{a}+\frac{r^{2}}{2}\alpha^{\prime}\left(\frac{b^{\prime}}{a}+4\frac{b}{a}\phi^{\prime}\right),\\
R_{rr}&=&\frac{3a^{\prime 2}}{4a^{2}}-\frac{b^{\prime 2}}{2b^{2}}+a\tilde{\Lambda}^{\prime}+\frac{1}{2}a^{\prime}\tilde{\Lambda}
-4\phi^{\prime\prime}+2\phi^{\prime}\left(\frac{a^{\prime}}{a}-\frac{b^{\prime}}{b}\right)\nonumber\\
&+&\frac{1}{r}\left(-4\phi^{\prime}-\frac{a^{\prime}+2b^{\prime}}{b}+\frac{2ab^{\prime}}{b^{2}}\right)-\frac{a^{\prime\prime}}{2a}+\frac{2(a-b)}{r^{2}b},\\
R_{\theta\theta}&=&\frac{r^{2}b}{a}\left(\frac{a^{\prime}}{a}\phi^{\prime}-2\phi^{\prime\prime}-4\phi^{\prime 2}\right)+\frac{r^{2}}{a}\left(\frac{b^{\prime 2}}{2b}-3b^{\prime}\phi^{\prime}-\frac{1}{2}b^{\prime\prime}\right)\nonumber\\
&+&\frac{1}{2}b^{\prime}\tilde{\Lambda}r^{2}+r\left(b\tilde{\Lambda}-\frac{b^{\prime}}{b}-6\frac{b}{a}\phi^{\prime}\right)+\frac{b}{a}-1.
\end{eqnarray}
%

%%%%%%%%%%%%%%%%%%%%%%%%%%%%%%%%%%%%%%%%%%%%%%%%

\end{document}